\documentclass[twocolumn]{aastex631}

\usepackage[english]{babel}
\usepackage{amsmath}
\usepackage{amssymb}
\usepackage{graphicx}
\usepackage{subfigure}
\usepackage[normalem]{ulem}
\usepackage{enumerate}
\usepackage{booktabs}

\usepackage{verbatim}

\usepackage{color}
\usepackage{multirow}
\usepackage{mathtools}
\usepackage{epstopdf}
\usepackage{tabularx}

\usepackage{url}
\usepackage{xcolor}

\def\apjl{ApJL}
\def\apjs{ApJS}
\def\aap{A\&A}

\def\be{\begin{equation}} 
\def\ee{\end{equation}} 
\def\ba{\begin{eqnarray}} 
\def\ea{\end{eqnarray}}

\def\HI{\hbox{H~$\scriptstyle\rm I\ $}}

\def\CIV{\hbox{C~$\scriptstyle\rm IV\ $}}

\def\gsim{\lower.5ex\hbox{\gtsima}} 
\def\lsim{\lower.5ex\hbox{\ltsima}} \def\gtsima{$\; \buildrel > \over 
\sim \;$} \def\ltsima{$\; \buildrel < \over \sim \;$} \def\prosima{$\; 
\buildrel \propto \over \sim \;$} \def\gsim{\lower.5ex\hbox{\gtsima}} 
\def\lsim{\lower.5ex\hbox{\ltsima}} 
\def\simgt{\lower.5ex\hbox{\gtsima}} 
\def\simlt{\lower.5ex\hbox{\ltsima}} 
\def\simpr{\lower.5ex\hbox{\prosima}}   
  
 \def\gtsima{$\; \buildrel > \over \sim \;$} 
\def\ltsima{$\; \buildrel < \over \sim \;$} 
\def\gsim{\lower.5ex\hbox{\gtsima}} 
\def\lsim{\lower.5ex\hbox{\ltsima}} 
\def\simgt{\lower.5ex\hbox{\gtsima}} 
\def\simlt{\lower.5ex\hbox{\ltsima}} 
\def\simpr{\lower.5ex\hbox{\prosima}}

\def\E3{{\cal E}_{\rm g}^{III}}

\def\r12{r_{1/2}} 
\def\x12{x_{1/2}} 
\def\v12{v_{1/2}}

\def\HI{\hbox{H~$\scriptstyle\rm I $~}}

\def\MGII{\hbox{Mg~$\scriptstyle\rm II $}} 
\def\CIV{\hbox{C~$\scriptstyle\rm IV $}}

\def\SII{\hbox{Si~$\scriptstyle\rm II $}} 
 
\def\OIa{\hbox{O~$\scriptstyle\rm I $}} 
\def\CIV{\hbox{C~$\scriptstyle\rm IV $}} 
\def\FeII{\hbox{Fe~$\scriptstyle\rm II $}} 
\def\AlII{\hbox{Al~$\scriptstyle\rm II $}}

\def\nh2{n_{\rm H2}}
\def\fh2{f_{\rm H2}}

\def\angstrom{\textrm{A\kern -1.3ex\raisebox{0.6ex}{$^\circ$}}}

\makeatletter
\def\@hex@@Hex#1%
 {\if a#1A\else \if b#1B\else \if c#1C\else \if d#1D\else
  \if e#1E\else \if f#1F\else #1\fi\fi\fi\fi\fi\fi \@hex@Hex}
\makeatother

\definecolor{apcolor}{HTML}{b3003b}
\definecolor{cbcolor}{HTML}{ff0f00}
\definecolor{afcolor}{HTML}{b3443c}
\definecolor{vgcolor}{HTML}{8F00FF}
\definecolor{tbdcolor}{HTML}{E8A95E}
\definecolor{stefcolor}{HTML}{0047ab}

\shorttitle{Chemical diagnostics for first stars' enrichment}
\shortauthors{Vanni, Salvadori, D'Odorico et al.}

\begin{document}

\title{Chemical Diagnostics to Unveil Environments Enriched by First Stars}

\correspondingauthor{Irene Vanni}
\email{irene.vanni@unifi.it}

\author[0000-0001-9647-0493]{Irene Vanni}
\author[0000-0001-7298-2478]{Stefania Salvadori}
\affiliation{Dipartimento di Fisica e Astrofisica, Universit\`a degli Studi di Firenze, Via G. Sansone 1, I-50019 Sesto Fiorentino, Italy}
\affiliation{INAF/Osservatorio Astrofisico di Arcetri, Largo E. Fermi 5, I-50125 Firenze, Italy}
\author[0000-0003-3693-3091]{Valentina D'Odorico}
\affiliation{INAF/Osservatorio Astronomico di Trieste, Via G. Tiepolo 11, I-34143, Trieste, Italy}
\affiliation{Scuola Normale Superiore, Piazza dei Cavalieri 7, I-56126, Pisa, Italy}
\affiliation{IFPU - Institute for Fundamental Physics of the Universe, via Beirut 2, I-34151 Trieste, Italy}
\author[0000-0003-2344-263X]{George D. Becker}
\affiliation{Department of Physics \& Astronomy, University of California, Riverside, CA-92521, USA}
\author[0000-0002-6830-9093]{Guido Cupani}
\affiliation{INAF/Osservatorio Astronomico di Trieste, Via G. Tiepolo 11, I-34143, Trieste, Italy}

\begin{abstract}
Unveiling the chemical fingerprints of the first (Pop~III) stars is crucial for indirectly studying their properties and probing their massive nature. In particular, very massive Pop~III stars explode as energetic Pair-Instability Supernovae (PISNe), allowing their chemical products to escape in the diffuse medium around galaxies, opening the possibility to observe their fingerprints in distant gas clouds. Recently, three $z>6.3$ absorbers with abundances consistent with an enrichment from PISNe have been observed with JWST. In this Letter, we present novel chemical diagnostics to uncover environments mainly imprinted by PISNe. Furthermore, we revise the JWST low-resolution measurements by analysing the publicly available high-resolution X-Shooter spectra for two of these systems. 
Our results reconcile the chemical abundances of these absorbers with those from literature, which are found to be consistent with an enrichment dominated ($> 50\%$ metals) by normal Pop~II SNe.
We show the power of our novel diagnostics in isolating environments uniquely enriched by PISNe from those mainly polluted by other Pop~III and Pop~II SNe. When the subsequent enrichment from Pop~II SNe is included, however, we find that the abundances of PISN-dominated environments
partially overlap with those predominantly enriched by other Pop~III and Pop~II SNe. We dub these areas \emph{confusion regions}.
Yet, the odd-even abundance ratios [Mg,Si/Al] are extremely effective in pinpointing PISN-dominated environments and allowed us to uncover, for the first time, an absorber consistent with a combined enrichment by a PISN and another Pop III SN for all the six measured elements.

\end{abstract}


\section{Introduction}
\label{sec:intro}

Understanding the properties of the first (Pop~III) stars and their impact on subsequent galaxy formation is a fundamental problem 
\citep[e.g.][]{Klessen2023}. Discovering the chemical fingerprints of Pop~III stars exploding as supernovae (SNe) is crucial to achieve this goal \citep[e.g.][]{Koutsouridou2023a}.
Indeed, cosmological simulations show that Pop~III stars were more massive than present-day stars \citep[e.g.][]{Hirano2014,Susa2014}, 
an idea also supported by the persistent lack of metal-free stars \citep[e.g.][]{Rossi2021}. Thus, most Pop~III stars might have rapidly 
disappeared as SNe, promptly enriching the pristine gas with their newly formed chemical products. 

Pop~III SNe with various progenitor masses and explosion energies synthesize different heavy elements \cite[e.g.][]{Heger2010,Kobayashi2012} 
thus producing distinctive chemical fingerprints \citep[see][]{vanni2023}. These unique signatures can be observed 
in distant gaseous absorbers directly enriched by these pristine sources \citep{Saccardi2023}, 
or in long-lived metal-poor stars born in Pop~III enriched environments \citep[e.g.][]{Frebel2019,Skuladottir2021}.
Stellar evolution calculations for the chemical products of very massive pair instability SNe (PISNe, $\rm M_* = 140 - 260 M_{\odot}$, \citealp{Barkat1967}) are 
extremely robust and predict unique abundance patterns, featuring a strong odd-even effect \citep{Heger2002a,Takahashi2018b}.

Many efforts have been dedicated to seek out these unique PISN fingerprints
\citep[see][]{Salvadori2019,Aguado2023b,Caffau2023} and finally one possible pure PISN descendant 
has been found in the LAMOST survey \citep{Xing2023}.   
If predominantly enriched by PISN \citep[see][]{Jeena2024}, this unique star will provide us 
with the first tight constraints on the initial mass function (IMF) of Pop~III stars \citep[][]{Koutsouridou2024}.

PISNe are very energetic ($\rm E_{SN} \in [10^{52}; 10^{53}]$~erg), thus their metals can easily escape their hosting galaxies \citep[e.g.][]{Bromm2001,Smith2015} filling the diffuse circum-galactic and inter-galactic media \citep[e.g.][]{Pallottini2014}.
Thus, the long-waited discovery of this PISN descendant opens the concrete possibility to 
pinpoint the chemical fingerprint of PISNe in high-$z$ diffuse absorbers too.
Very recently, \citet{Christensen2023} reported chemical abundances consistent 
with a PISN enrichment for three $z>6.3$ absorbers observed in JWST QSO's spectra.
Indeed, they found that these systems have high [Si/O]$>0.5$ values and only upper limits for [C/O]$<0$ 
\citep[see e.g.][]{Ma2017b, Vale2024}. 

In this letter, we aim at addressing the reliability of the JWST low-resolution measurements and the determination of the key abundance ratios that need to be measured in order to identify PISN-enriched environments. Our final goal is to provide new chemical diagnostics to pinpoint distant absorbers predominantly enriched by PISNe and other Pop~III SNe. 

\section{Summary of the model}

\label{sec:model}

For this study, we use the model first introduced by \citet{Salvadori2019} and then generalised by \citet{Vanni2023b}.
This simple and general parametric study investigates the chemical abundances (elements from C to Zn) of 
gaseous environments imprinted by a single Pop~III SN and subsequently polluted by normal (Pop~II) SNe. Pop~III stars are indeed expected to predominantly form in isolation in low-mass 
mini-halos \citep[e.g.][]{Hirano2015}. Furthermore, simulations show that in these proto-galaxies self-enrichment from a single Pop~III SN can promptly 
trigger the formation of normal Pop~II stars by enriching the inter-stellar medium (ISM) above the critical metallicity value \citep[e.g.][]{Ritter2012}. The model is quite robust
since our findings for ancient metal-poor stars \citep{Vanni2023b, Skuladottir2023} have been confirmed by more sophisticated cosmological chemical evolution models and simulations \citep{Rossi2023, Koutsouridou2023a}. The model was used to unveil the Pop~III star's signature in distant gaseous absorbers \citep[][Sodini et al. submitted]{Salvadori2023}, which is a proof of its versatility.

The innovative aspects of our approach with respect to other simple models available in literature \citep[e.g.][]{Ishigaki2018,Welsh2019} are reported below.\\

{\bf A parametric study.} The uncertainties related to early cosmic star-formation are enclosed in two free parameters: 
the star-formation efficiency, $f_*$, which quantifies the amount of cold gas turned into stars, and the dilution factor, 
$f_{\rm dil}$, which varies depending upon the amount of gas available to dilute metals. A third parameter, $f_{\rm PopIII}$, 
sets the fraction of metals in the ISM provided by Pop~III with respect to Pop~II SNe. 
The chemical abundances of the gas, [X/H], depend on the ratio $f_*/f_{\rm dil}$ and $f_{\rm PopIII}$ \citep[see][]{Vanni2023b}, 
and thus are evaluated by varying these unknowns in the full parameter space: $f_*/f_{\rm dil}\in [10^{-4}; 10^{-1}]$ and $f_{\rm PopIII}\in [1,0.01]$. Conversely, the relative abundance ratios of different chemical elements, [X/Y], are strongly influenced by $f_{\rm PopIII}$ but they only indirectly depend on $f_*/f_{\rm dil}$, which sets the initial metallicity of subsequent generations of Pop~II stars \citep[see][]{Salvadori2019}.\\ 

{\bf Exploring the properties of Pop~III SNe.} 
To investigate the whole range of unknown properties of Pop~III SNe, we exploit the tabulated yields 
of \citet{Heger2010}, which represent the most complete datasets for Pop~III SNe with progenitor masses between $\rm 10-100~ M_{\odot}$.  
Indeed, the authors investigate the chemical products of these SNe, that can potentially explode with different energies: from the lowest-energy 
\emph{faint SNe}, $\rm E_{SN}=0.6 \times 10^{51}$~erg, to the most energetic \emph{hypernovae}, $\rm E_{SN}=10^{52}$~erg, adopting different mixing 
parameters. Here we explore this full parameter space. For more massive, $\rm m_{\star}=140-260 M_{\odot}$, Pop~III stars exploding as energetic PISNe, 
$\rm E_{SN}=10^{52}-10^{53}$~erg, we adopt the yields by \citet{Heger2002a}.
For our exploration we are thus implicitly assuming that Pop~III SNe with different progenitor mass within the range $\rm m_{\star}=10-100 M_{\odot}$
and $\rm m_{\star}=140-260 M_{\odot}$, and different SN explosion energies have an equal probability to explode. 
Note that considering a burst of Pop~III SNe with the same mass/explosion energy instead of a single SN would not change the results. Finally, here we 
also investigate how the abundance ratios of PISNe-enriched environments change after the explosion of a second Pop~III SN (see Appendix~\ref{app:twopopIII}).\\

{\bf Subsequent pollution by Pop~II SNe.} To study how the chemical abundances vary due to the further contribution of Pop~II SNe, 
we assume that Pop~II stars are distributed according to a Larson's IMF, $\phi (m_\star) \propto m_\star^{-2.35} \times \exp{\left(-{m_{\rm ch}}/{m_\star}\right)}$ with $m_{\rm ch}=0.35\rm M_{\odot}$ and compute the contribution of Pop~II stars above different masses, which allow us to evaluate the effect of SNe 
evolving on different time-scales \citep[see][]{Salvadori2019}. For Pop~II SNe, we adopt the recent yields from \citet[][non-rotating, set R]{Limongi2018}. \\ 

\section{Observational data}

\begin{table*}[t]
    \centering
    \begin{tabular}{c c c c }
    \hline
    QSO & UHS J0439+1634 & ULAS J1342+0928 & ULAS J1342+0928 \\         
    & $z_{\rm abs}=6.2897$ & $z_{\rm abs}=7.36899$ & $z_{\rm abs}=7.44345$ \\
    \hline
    [Si/O] & $>-0.46$ & $0.10\pm0.21$ & $-0.03\pm0.14$ \\
    $\mathrm{[C/O]}$ &  $>-0.57$  & $-0.20\pm0.30$ & $-0.58\pm0.34$ \\
    $\mathrm{[C/Fe]}$ & $0.55\pm0.09$ & $>-0.06$ & $>-0.96$ \\
    $\mathrm{[O/Fe]}$ & $<1.12$ & $>0.14$ & $>-0.38$ \\
    $\mathrm{[Si/Fe]}$ & $0.66\pm0.05$ & $>0.24$ & $>-0.41$ \\
    $\mathrm{[Mg/Fe]}$ & $0.32\pm0.04$ & -- &  -- \\
    $\mathrm{[Mg/Si]}$ & $-034\pm0.05$ & -- & -- \\
    $\mathrm{[Al/Si]}$ & $<-0.83$ & -- & -- \\
    $\mathrm{[Al/Mg]}$ & $<-0.49$ & -- & --\\ 
    \hline
    \end{tabular}
    \caption{Revised abundance ratios for the systems reported by  \citet[][]{Christensen2023} along the sightlines to UHS J0439+1634 and ULAS J1342+0928 based on the X-Shooter spectra.}
    \label{tab:obs}
\end{table*}

\label{sec:data}

Our compiled literature sample consists of 50 different metal-poor absorption systems analysed in \citet[][]{Cooke2011}, \citet{Becker2012}, \citet{Welsh2022}, \citet{Saccardi2023} (only with [Fe/H]$\leq -1$) and \citet{Christensen2023}. The redshifts of the absorbers range between $\sim 2$ \citep[][]{Cooke2011} and $\sim 7.5$ \citep[][]{Christensen2023}. These absorbers have different nature: 24 are Damped Lyman-$\alpha$ systems (DLAs, \citealp{Cooke2011,Welsh2022}) and 19 more diffuse sub-DLAs and Lyman Limit systems (LLS, \citealp{Saccardi2023}). 
The systems at $z\gtrsim5$ \citep[][]{Becker2012,Christensen2023} were selected from the presence of the \OIa\ absorption, tracer of neutral gas, because the \HI content is hardly measurable at these redshifts. Therefore, they are DLAs or sub-DLAs, but their precise nature is unknown. The main elements whose abundances can be measured are C, O, Mg, Al, Si, and Fe, but these elements are simultaneously available for 9 absorbers only.

\subsection{New measurements}
The analysis of the three systems with possible PISN enrichment by \citet{Christensen2023} have been carried out in JWST/NIRSpec spectra at low spectral resolution ($\rm R\approx 2700$). Two of these systems fall in the spectra of the QSOs 
UHS J0439+1634 ($z_{\rm em} = 6.5185$) and ULAS J1342+0928 ($z_{\rm em} = 7.5353$), respectively, which also have intermediate resolution ($\rm R\approx 9000$) good signal-to-noise ratio VLT/X-Shooter data. 

The X-Shooter spectrum of J0439+1634 was presented in \citet{Dodorico2023} and analysed in \citet{RDavies2023a}. However, these authors do not identify the \OIa\ 1302 $\mathring{\rm A}$ transition associated with the system at $z=6.2897$, discussed by \citet{Christensen2023}, because it falls in a spectral region already populated by other absorption lines \citep[see][for details]{RDavies2023a}. Our analysis of the system is reported in Appendix~\ref{app:abs_sys}.

The analysis of the X-shooter absorption spectrum of J1342+0928 is carried out in this work for the first time, and is described in Appendix~\ref{app:abs_sys}. 

For the purpose of this paper, we analysed only the two \OIa\ absorption systems at $z=7.369$ and $z=7.443$ for which  \citet{Christensen2023} reported the chemical abundances, the former showing possible PISN enrichment. However, we carried out a first look analysis of the whole spectrum to verify that the transitions of the considered systems were not blended with lines from other systems. 

Observed absorption features were identified and fitted with Voigt profiles to determine ionic column densities. 
We derived chemical abundances from the ratio of the measured column densities and adopting solar chemical abundances from \citet{Asplund2009a}. Based on the common hypothesis that these systems are metal-poor DLAs \citep[e.g.][]{Becker2012,Christensen2023}, we did not correct for ionization effects and dust depletion. Measured column densities and plots of the absorption lines are reported in Appendix~\ref{app:abs_sys}. 

\begin{figure*}[h]
    \centering
    \includegraphics[width=1\linewidth]{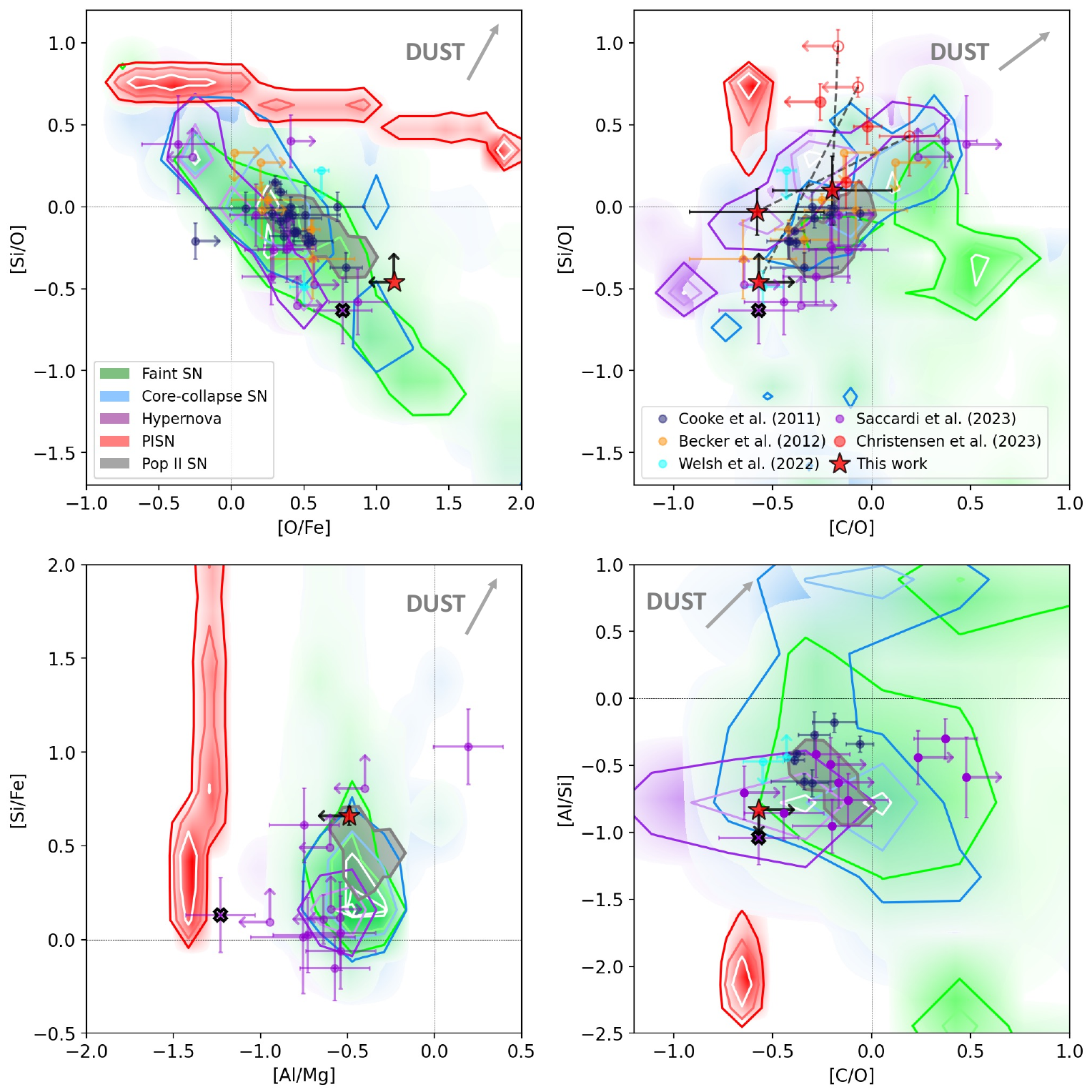}
    \caption{Chemical abundances predicted for environments uniquely enriched by PISNe (red), by other Pop~III SNe (faint SNe: green; core-collapse SNe: blue; hypernovae: purple), and by normal Pop~II SNe (grey shaded areas). Each environment is enriched by a single Pop~III SN with different mass and explosion energy. The contours with decreasing color intensity identify 30\%, 60\% and 
    90\% probability densities, corresponding to the fraction of environments predicting those abundances for each SN type. 
    Points with errorbars are the measured abundances of absorbers in the literature (Sec.~\ref{sec:data}, labels). The values reported by \citet[red empty]{Christensen2023} are connected with the new measurements obtained using higher-resolution spectra (star). The absorber in the sightline to J1018+0548 \citep[][$z = 3.39$]{Saccardi2023} is highlighted with a black cross (see Sec.~\ref{sec:PopII_1}). The arrows show in which direction dust depletion would modify the predicted abundances.}
    \label{fig:PopIII_only}
\end{figure*}

The abundances for the system in UHS J0439+1634 and the 2 systems in ULAS J1342+0928  as derived by \citet{Christensen2023} are shown in Fig.~\ref{fig:PopIII_only} with red empty points. These points are connected with dashed black lines to the revised abundances derived in this work (red stars), which are reported in Table~\ref{tab:obs}. The new abundances derived from X-Shooter data agree with other systems in the literature, thanks to a decrease of the [Si/O] values of these absorbers of $\gtrsim 0.4$~dex. 
The changes in relative abundances are generally due to an overestimate of the \SII\ column densities in the NIRSpec spectra, possibly deriving from line blending. The only point from \citet{Christensen2023} that remains consistent with an enrichment from PISNe is due to the system at $z=6.5625$ toward VDES J0020-3653, which at present does not have sufficient high-resolution observations to carry out the analysis of absorption lines.

\section{Chemical diagnostics}

In this Section, we introduce novel diagnostics to pinpoint systems predominantly enriched by PISNe.

\subsection{Purely Pop~III enriched environments}

\label{sec:PopIII}

In Fig.~\ref{fig:PopIII_only}, we compare the chemical abundance ratios measured in metal-poor absorbers with the predictions of our model for environments {\it uniquely} polluted by Pop~III SNe, i.e., assuming $f_{\rm PopIII}=1.0$ and varying the mass and explosion energy of Pop~III SNe (see Sec.~\ref{sec:model}). 
For reference we also show the predictions for environments polluted by normal Pop~II SNe ($f_{\rm PopIII}=0.01$).

The {\it chemical diagnostics} presented here were specifically selected to discern the environments enriched by PISNe from other SN types. Indeed, we see that PISN-polluted environments reside in very specific regions of the plots, having quite high [Si/O]$> 0.2$, low values of [C/O]$<-0.5$ and [Al/Si]$<-1.5$, and the lowest values of [Al/Mg]$< -1.0$. Thus, the combination of the abundance ratios shown in Fig.~\ref{fig:PopIII_only} represents a powerful tool to pinpoint PISN-enriched environments. 
On the contrary, the environments enriched by the other types of Pop~III SNe are not completely separated one from the other and their chemical abundance ratios partially overlap with those imprinted by normal Pop~II SNe.

In Fig.~\ref{fig:PopIII_only} (top right), we see that only the system toward VDES J0020-3653 in \citet{Christensen2023} remains consistent with an imprint from PISN.
Comparing the [Si/O] value of this absorber with the predictions of our model for the environments polluted solely by PISN, we determine the possible initial masses for the PISN that enriched this absorber, $m_* \in [183.0; 234.0] \rm M_{\odot}$. Then, we predict the range of abundances for other elements, which would imply that this absorber was truly and uniquely polluted by PISNe: [O/Fe]$\simeq[-0.35;1.04]$, [Al/Si]$\simeq[-2.16;-1.95]$, [Si/Fe]$\simeq [0.40;1.58]$ and [Al/Mg]$\simeq[-1.38;-1.28]$.

Finally, we see that the different chemical abundances predicted for Pop~III enriched environments and measured in various absorbers nicely follow the same trends, with several systems overlapping with the predicted abundances in the various chemical diagnostics. Does this mean that all these absorbers have been imprinted by Pop~III SNe?

\subsection{Pop~II contribution: the confusion region}
\label{sec:PopII_1}

\begin{figure*}
    \centering
    \includegraphics[width=1\linewidth]{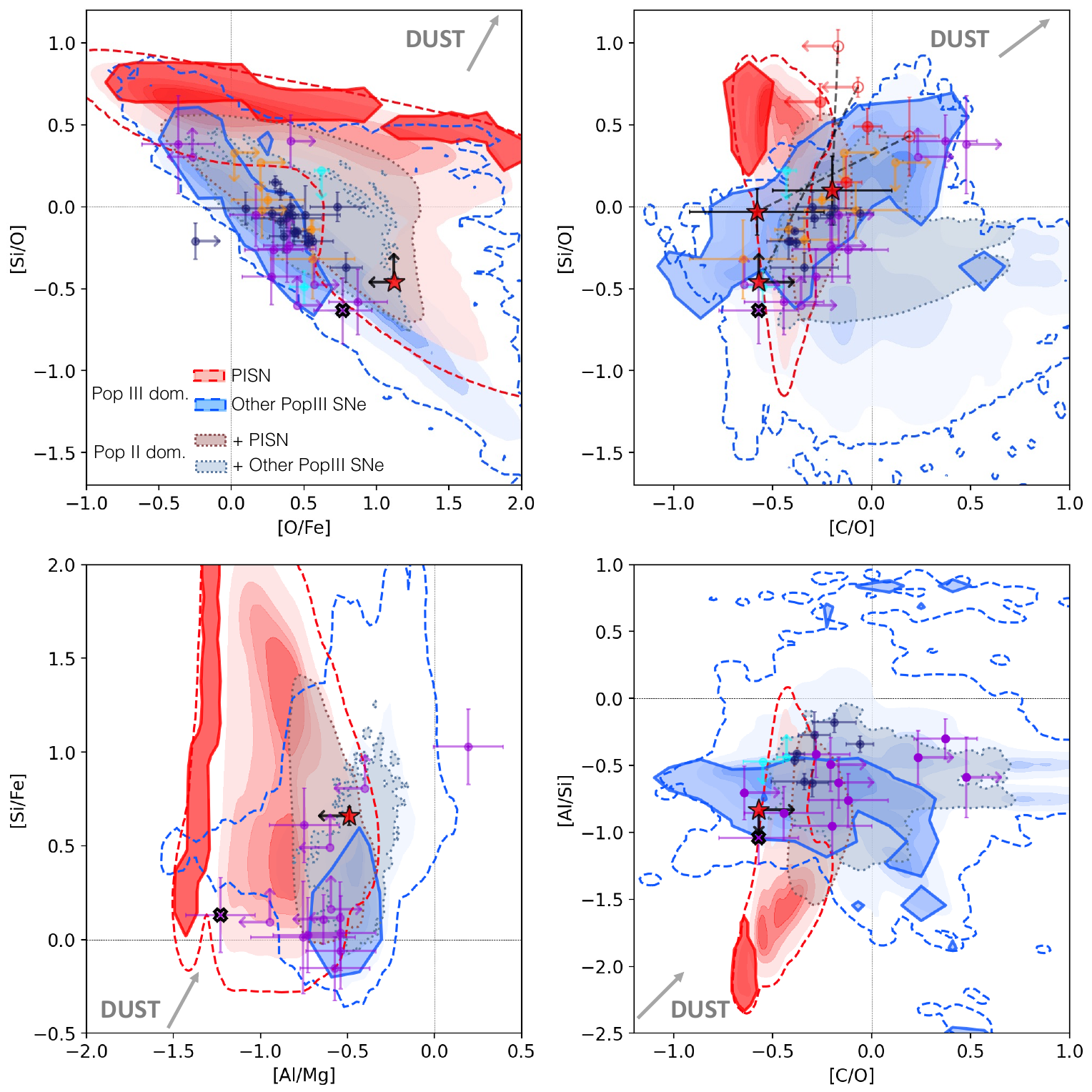}
    \caption{The same as Fig.~\ref{fig:PopIII_only} but for environments whose metals have been inherited: uniquely by PISNe/other Pop~III SNe (red/blue solid line, $f_{\rm PopIII}=$1); mainly by PISNe/other Pop~III SNe but also by normal Pop~II SNe (red/blue dashed lines, $f_{\rm PopIII}\geq$ 0.6); mainly by normal Pop~II SNe but also by PISNe/other Pop~III SNe (brown/gray dotted lines, $f_{\rm PopIII}\leq $0.5).}
    \label{fig:Maps_DLAs}
\end{figure*}

In Fig.~\ref{fig:Maps_DLAs} we show how the injection of heavy elements from Pop~II SNe changes the chemical 
abundances of PISN and other Pop~III SN pre-enriched environments. To this end, we compute the probability density regions
of environments solely ($100\%$ metals), mainly ($\geq 60\%$), and only partially ($\leq 50\%$) enriched by 
Pop~III (PISNe or other SNe), thus varying $f_{\rm PopIII}$. We see that the contribution of normal Pop~II SNe 
enlarges the regions predominantly imprinted by PISNe, which partially overlap with those 
mainly enriched by other Pop~III/II SNe. Furthermore, while there is still a partial separation between 
environments mainly polluted by PISNe and by other Pop~III SNe, when Pop~II SNe become the dominant contributors 
the two regions almost completely overlap, meaning that we are loosing the information about the pristine SNe 
that initially enriched these media.

In each diagnostic of Fig.~\ref{fig:Maps_DLAs}, we notice the emergence of regions in which the chemical abundance 
ratios predicted for Pop~III- (PISNe and other pristine SNe) and Pop~II-dominated environments blend. 
We dub this area the \emph{confusion region}, since environments dwelling in this specific zone might 
have been imprinted by different sources: one Pop~III SN (or two, see Appendix~\ref{app:twopopIII}), 
a mixture of Pop~III and Pop~II SNe or a population of normal Pop~II SNe. 
In Fig.~\ref{fig:Maps_DLAs}, we see that the majority of the absorption systems present in the literature 
reside within the confusion regions and in particular, in most of the proposed diagnostics, they are 
consistent with an enrichment dominated by normal Pop~II SNe, but not solely driven by Pop~II SNe (see Fig.~\ref{fig:PopIII_only}). 
Thus, the nature of the absorbers dwelling in the confusion regions might remain undetermined, even 
by studying their abundance patterns with the currently available $\leq 5$ abundance ratios.

\begin{figure*} [h] \centering\includegraphics[width=1\linewidth]{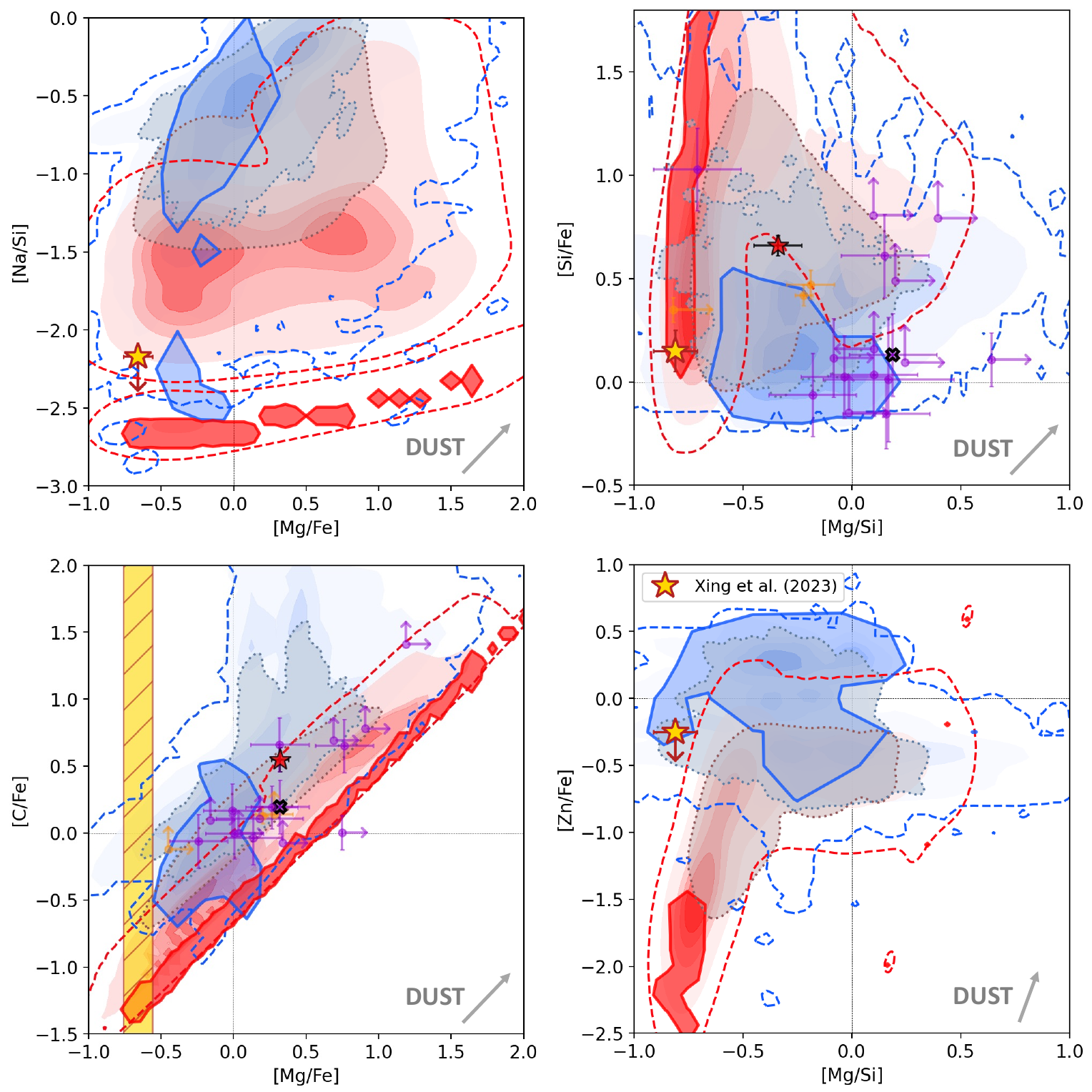}
    \caption{The same as Fig.~\ref{fig:Maps_DLAs}. The yellow star and shaded area are the abundances of \citeauthor{Xing2023}'s \citeyear{Xing2023} star: the possible pure PISN descendant.}
    \label{fig:Maps_PISN}
\end{figure*}

To pinpoint PISN- and other Pop~III-enriched environments, we should thus look for absorbers that 
dwell outside of the confusion regions, i.e. in areas where the Pop~III chemical fingerprints are
preserved. To this end, we see that the odd-even abundance ratios are extremely effective in locating 
PISN dominated environments, for example [Al/Mg]~$< -0.75$ 
and [Al/Si]~$<-1.5$ \citep[see also][]{Takahashi2018}. The absorption system at $z=3.39$ in the spectrum of J1018+0548 
from \citet[][black cross in Figs.~\ref{fig:PopIII_only},\ref{fig:Maps_DLAs},\ref{fig:Maps_PISN}]{Saccardi2023} lies at [Al/Mg]~$=-1.23$ and [Si/Fe]~$=0.13$ implying that it might be 
predominantly enriched by PISNe. However, when we consider its positions in other diagnostic plots, 
we see that this absorber is also consistent with being predominantly enriched by both 
PISNe and other Pop III SNe. 

So far, we have confined our discussion to elements commonly measured in high-$z$ absorbers. However, in nearby stars, we typically measure other elements. For example, for the newly discovered LAMOST star likely imprinted by PISNe \citep[J1010+2358,][]{Xing2023}, it was not possible to measure the abundances of C, O and Al, while they were determined for many elements heavier than Ca.  
In Fig.~\ref{fig:Maps_PISN}, we show other four chemical diagnostics for which the star J1010+2358 is always consistent with a PISN dominated enrichment. 
Measuring the C (or the Zn) abundance would be fundamental, since a value of [C/Fe]$<-1.2$ (or [Zn/Fe]$<-1.5$) would imply an exclusive enrichment by PISNe.

\subsection{A unique absorber}

\label{sec:unique_abs}
\begin{figure}
    \centering
    \includegraphics[width=\linewidth]{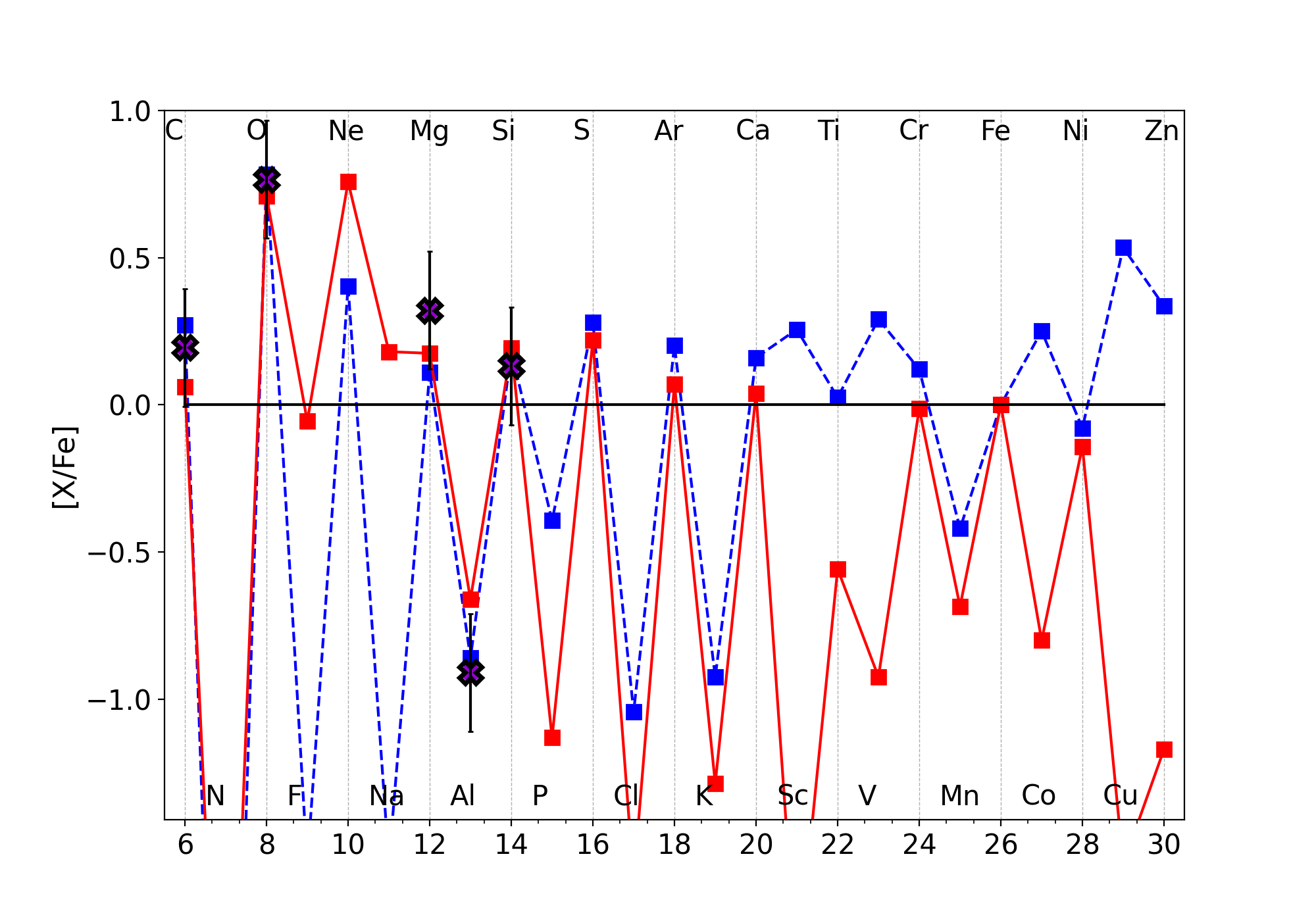}
    \caption{Abundance pattern of the absorber J1018+0548 (\citealp{Saccardi2023}, black crosses) 
    compared with the two best fitting models predicting an enrichment from: (i) a PISN and a 
    Pop~III SN (red); (ii) two Pop~III SNe (blue).}
    \label{fig:chi_square}
\end{figure}

The absorber J1018+0548 agrees with an enrichment dominated by PISN or other Pop~III SNe in all the proposed diagnostics (Fig.~\ref{fig:Maps_DLAs}-\ref{fig:Maps_PISN}). By applying a chi square diagnostics (see Appendix~\ref{app:chisq}), we looked for the 
models that best reproduce its abundance pattern, i.e. with reduced chi square, $\chi^2_{\nu}$, lower than two times the minimum one (Fig.~\ref{fig:chi_square}). The minimum 
$\chi^2_{\nu}=0.19$ corresponds to an enrichment driven by two Pop~III SNe with equal explosion energy 
($\rm E_{SN}=1.8\times 10^{51}$~erg) but very different masses ($\rm \approx 10$ and $100 \rm M_{\odot}$, see Table~\ref{tab:bestmod}). 
However, a low $\chi^2_{\nu}=$~0.37, is also obtained when considering a pollution from a $\rm 253 M_{\odot}$ PISN 
(contributing to 10\% of the metals), and a $\rm \approx 30 M_{\odot}$ Pop~III SN with $\rm E_{SN} = 1.2 \times 10^{51}$~erg. 
Both models fit very well the abundance ratios of the five light elements measured in the J1018+0548 absorber. 
However they strongly differ for elements heavier than Ca, in particular [Zn/Fe] which is $\approx 1.5$~dex 
lower for massive PISN enrichment.
\\

\section{Conclusions}

\label{sec:Conclusions}

In this Letter, we present novel chemical diagnostics to isolate PISN-enriched environments
from those imprinted by other Pop~III SNe and normal Pop~II stars. To this end, we exploited 
the simple and general parametric model by \cite{Vanni2023b}. We show that environments enriched 
by a single PISN dwell in well defined zones of the proposed chemical diagnostics. Indeed they 
cover $>2$ dex in [X/Fe] but have characteristic values of [Al, Mg, Na/Si] and [Si, C/O], 
covering $<0.5$ dex. 
However, when the contribution of normal Pop~II SNe is accounted for, regions predominantly 
imprinted  by PISNe ($\geq 60\%$ of metals) grow and blend with those mainly enriched by 
other Pop~III or Pop~II SNe, creating the so-called {\it confusion regions}. 
Our predictions are very general and thus can be used to interpret the chemical properties 
of various environments across cosmic times: high-z emitting galaxies \citep[e.g.][]{DEugenio2023}, 
distant gaseous absorbers in the sightline of QSOs or GRBs \citep[see also][]{Ma2017} and
ancient stellar fossils in our Local Group \citep{Vanni2023b}.

By comparing our model predictions with measurements in high-$z$ absorbers, we see that most systems in the 
literature have elemental abundances consistent with being mainly imprinted by normal Pop~II SNe, in agreement 
with previous studies \citep[e.g.][]{Becker2012,Salvadori2012}. Note that we do not account for dust depletion 
since most absorbers in our literature sample are metal-poor, meaning that dust should not 
affect substantially their chemical abundances \citep{Vladilo2018, Saccardi2023}. 
In Figs.~\ref{fig:PopIII_only}, \ref{fig:Maps_DLAs}, \ref{fig:Maps_PISN}, we show in which direction the predictions
would shift if dust was present, using the prescriptions of \citet{Konstantopoulou2023} \citep[see also][]{DeCia2016}.  

Our analysis of the publicly available high-resolution X-Shooter spectra for two QSOs presented by \cite{Christensen2023}, reconcile the chemical abundances of two absorption systems, suggesting exclusive PISN-enrichment, with literature data, i.e., with an enrichment dominated 
by normal Pop~II SNe. Only the system at $z=6.5625$ along the sightline to VDES J0020–3653, for which high resolution 
spectra are not available, remains consistent with a PISN pollution for the {\it three} measured elements. Conversely, 
the Galactic halo star observed at high-resolution by \citet[][12 elements]{Xing2023} is in agreement with an exclusive 
PISN-enrichment in {\it all} considered diagnostics. Still, more chemical elements (e.g., C) need to be measured to confirm 
a pure pollution by a massive PISN \citep[see also][]{Koutsouridou2024}.

Our novel chemical diagnostics allowed us to uncover an absorption system \citep[$z=3.39$ in J1018+0548,][]{Saccardi2023} 
that is consistent with being {\it uniquely} enriched by Pop~III SNe (see Fig.~\ref{fig:chi_square}). Notwithstanding the six chemical elements measured, to understand if this absorber was enriched by a PISN, it is crucial to measure zinc (Sec.~\ref{sec:unique_abs}). The high-resolution spectrograph ANDES for the ELT will allow us to achieve this goal, understanding the nature of this unique absorber \citep{Dodorico2023}.

Ultimately, armed with our novel chemical diagnostics and ANDES/ELT, we will be able to unveil these precious distant gas clouds, fully complementing stellar archaeology to shed light on the first stars and their role on early galaxy formation.

\label{sec:discuss}

\section*{Acknowledgements}
This project received funding from the ERC Starting Grant NEFERTITI H2020/808240 (PI: S. Salvadori). GDB was supported by the National Science Foundation through grant AST-1751404.
\bibliography{Biblio}

\begin{thebibliography}{}
\expandafter\ifx\csname natexlab\endcsname\relax\def\natexlab#1{#1}\fi
\providecommand{\url}[1]{\href{#1}{#1}}
\providecommand{\dodoi}[1]{doi:~\href{http://doi.org/#1}{\nolinkurl{#1}}}
\providecommand{\doeprint}[1]{\href{http://ascl.net/#1}{\nolinkurl{http://ascl.net/#1}}}
\providecommand{\doarXiv}[1]{\href{https://arxiv.org/abs/#1}{\nolinkurl{https://arxiv.org/abs/#1}}}

\bibitem[{{Aguado} {et~al.}(2023){Aguado}, {Salvadori}, {Sk{\'u}lad{\'o}ttir}, {Caffau}, {Bonifacio}, {Vanni}, {Gelli}, {Koutsouridou}, \& {Amarsi}}]{Aguado2023b}
{Aguado}, D.~S., {Salvadori}, S., {Sk{\'u}lad{\'o}ttir}, {\'A}., {et~al.} 2023, \mnras, 520, 866, \dodoi{10.1093/mnras/stad164}

\bibitem[{{Asplund} {et~al.}(2009){Asplund}, {Grevesse}, {Sauval}, \& {Scott}}]{Asplund2009a}
{Asplund}, M., {Grevesse}, N., {Sauval}, A.~J., \& {Scott}, P. 2009, \araa, 47, 481, \dodoi{10.1146/annurev.astro.46.060407.145222}

\bibitem[{{Barkat} {et~al.}(1967){Barkat}, {Rakavy}, \& {Sack}}]{Barkat1967}
{Barkat}, Z., {Rakavy}, G., \& {Sack}, N. 1967, \prl, 18, 379, \dodoi{10.1103/PhysRevLett.18.379}

\bibitem[{{Becker} {et~al.}(2012){Becker}, {Sargent}, {Rauch}, \& {Carswell}}]{Becker2012}
{Becker}, G.~D., {Sargent}, W. L.~W., {Rauch}, M., \& {Carswell}, R.~F. 2012, \apj, 744, 91, \dodoi{10.1088/0004-637X/744/2/91}

\bibitem[{{Bromm} {et~al.}(2001){Bromm}, {Ferrara}, {Coppi}, \& {Larson}}]{Bromm2001}
{Bromm}, V., {Ferrara}, A., {Coppi}, P.~S., \& {Larson}, R.~B. 2001, \mnras, 328, 969, \dodoi{10.1046/j.1365-8711.2001.04915.x}

\bibitem[{{Caffau} {et~al.}(2023){Caffau}, {Lombardo}, {Mashonkina}, {Sitnova}, {Bonifacio}, {Matas Pinto}, {Kordopatis}, {Sestito}, {Aguado}, {Salvadori}, {Spite}, {Fran{\c{c}}ois}, {Sbordone}, {Mucciarelli}, {Martin}, {Fern{\'a}ndez-Alvar}, \& {Gonz{\'a}lez Hern{\'a}ndez}}]{Caffau2023}
{Caffau}, E., {Lombardo}, L., {Mashonkina}, L., {et~al.} 2023, \mnras, 518, 3796, \dodoi{10.1093/mnras/stac3372}

\bibitem[{{Christensen} {et~al.}(2023){Christensen}, {Jakobsen}, {Willott}, {Arribas}, {Bunker}, {Charlot}, {Maiolino}, {Marshall}, {Perna}, \& {{\"U}bler}}]{Christensen2023}
{Christensen}, L., {Jakobsen}, P., {Willott}, C., {et~al.} 2023, \aap, 680, A82, \dodoi{10.1051/0004-6361/202347943}

\bibitem[{{Cooke} {et~al.}(2011){Cooke}, {Pettini}, {Steidel}, {Rudie}, \& {Nissen}}]{Cooke2011}
{Cooke}, R., {Pettini}, M., {Steidel}, C.~C., {Rudie}, G.~C., \& {Nissen}, P.~E. 2011, \mnras, 417, 1534, \dodoi{10.1111/j.1365-2966.2011.19365.x}

\bibitem[{{Cupani} {et~al.}(2022){Cupani}, {D'Odorico}, {Cristiani}, {Russo}, {Calderone}, \& {Taffoni}}]{Cupani2022}
{Cupani}, G., {D'Odorico}, V., {Cristiani}, S., {et~al.} 2022, in Astronomical Society of the Pacific Conference Series, Vol. 532, Astronomical Society of the Pacific Conference Series, ed. J.~E. {Ruiz}, F.~{Pierfedereci}, \& P.~{Teuben}, 207

\bibitem[{{Davies} {et~al.}(2023){Davies}, {Ryan-Weber}, {D'Odorico}, {Bosman}, {Meyer}, {Becker}, {Cupani}, {Bischetti}, {Sebastian}, {Eilers}, {Farina}, {Wang}, {Yang}, \& {Zhu}}]{RDavies2023a}
{Davies}, R.~L., {Ryan-Weber}, E., {D'Odorico}, V., {et~al.} 2023, \mnras, 521, 289, \dodoi{10.1093/mnras/stac3662}

\bibitem[{{De Cia} {et~al.}(2016){De Cia}, {Ledoux}, {Mattsson}, {Petitjean}, {Srianand}, {Gavignaud}, \& {Jenkins}}]{DeCia2016}
{De Cia}, A., {Ledoux}, C., {Mattsson}, L., {et~al.} 2016, \aap, 596, A97, \dodoi{10.1051/0004-6361/201527895}

\bibitem[{{D'Eugenio} {et~al.}(2023){D'Eugenio}, {Maiolino}, {Carniani}, {Curtis-Lake}, {Witstok}, {Chevallard}, {Charlot}, {Baker}, {Arribas}, {Boyett}, {Bunker}, {Curti}, {Eisenstein}, {Hainline}, {Ji}, {Johnson}, {Looser}, {Nakajima}, {Nelson}, {Rieke}, {Robertson}, {Scholtz}, {Smit}, {Venturi}, {Tacchella}, {Uebler}, {Willmer}, \& {Willott}}]{DEugenio2023}
{D'Eugenio}, F., {Maiolino}, R., {Carniani}, S., {et~al.} 2023, arXiv e-prints, arXiv:2311.09908, \dodoi{10.48550/arXiv.2311.09908}

\bibitem[{{D'Odorico} {et~al.}(2023{\natexlab{a}}){D'Odorico}, {Bolton}, {Christensen}, {De Cia}, {Zackrisson}, {Kordt}, {Izzo}, {Li}, {Maiolino}, {Marconi}, {Richter}, {Saccardi}, {Salvadori}, {Vanni}, {Feruglio}, {Fumagalli}, {Fynbo}, {Noterdaeme}, {Papaderos}, {Peroux}, {Verma}, {Di Marcantonio}, {Origlia}, \& {Zanutta}}]{Vale2024}
{D'Odorico}, V., {Bolton}, J.~S., {Christensen}, L., {et~al.} 2023{\natexlab{a}}, arXiv e-prints, arXiv:2311.16803, \dodoi{10.48550/arXiv.2311.16803}

\bibitem[{{D'Odorico} {et~al.}(2023{\natexlab{b}}){D'Odorico}, {Ba{\~n}ados}, {Becker}, {Bischetti}, {Bosman}, {Cupani}, {Davies}, {Farina}, {Ferrara}, {Feruglio}, {Mazzucchelli}, {Ryan-Weber}, {Schindler}, {Sodini}, {Venemans}, {Walter}, {Chen}, {Lai}, {Zhu}, {Bian}, {Campo}, {Carniani}, {Cristiani}, {Davies}, {Decarli}, {Drake}, {Eilers}, {Fan}, {Gaikwad}, {Gallerani}, {Greig}, {Haehnelt}, {Hennawi}, {Keating}, {Kulkarni}, {Mesinger}, {Meyer}, {Neeleman}, {Onoue}, {Pallottini}, {Qin}, {Rojas-Ruiz}, {Satyavolu}, {Sebastian}, {Tripodi}, {Wang}, {Wolfson}, {Yang}, \& {Zanchettin}}]{Dodorico2023}
{D'Odorico}, V., {Ba{\~n}ados}, E., {Becker}, G.~D., {et~al.} 2023{\natexlab{b}}, \mnras, 523, 1399, \dodoi{10.1093/mnras/stad1468}

\bibitem[{{Frebel} {et~al.}(2019){Frebel}, {Ji}, {Ezzeddine}, {Hansen}, {Chiti}, {Thompson}, \& {Merle}}]{Frebel2019}
{Frebel}, A., {Ji}, A.~P., {Ezzeddine}, R., {et~al.} 2019, \apj, 871, 146, \dodoi{10.3847/1538-4357/aae848}

\bibitem[{{Hartwig} {et~al.}(2018){Hartwig}, {Yoshida}, {Magg}, {Frebel}, {Glover}, {G{\'o}mez}, {Griffen}, {Ishigaki}, {Ji}, {Klessen}, {O'Shea}, \& {Tominaga}}]{Hartwig2018}
{Hartwig}, T., {Yoshida}, N., {Magg}, M., {et~al.} 2018, \mnras, 478, 1795, \dodoi{10.1093/mnras/sty1176}

\bibitem[{{Heger} \& {Woosley}(2002)}]{Heger2002a}
{Heger}, A., \& {Woosley}, S.~E. 2002, \apj, 567, 532, \dodoi{10.1086/338487}

\bibitem[{{Heger} \& {Woosley}(2010)}]{Heger2010}
---. 2010, \apj, 724, 341, \dodoi{10.1088/0004-637X/724/1/341}

\bibitem[{{Hirano} {et~al.}(2015){Hirano}, {Hosokawa}, {Yoshida}, {Omukai}, \& {Yorke}}]{Hirano2015}
{Hirano}, S., {Hosokawa}, T., {Yoshida}, N., {Omukai}, K., \& {Yorke}, H.~W. 2015, \mnras, 448, 568, \dodoi{10.1093/mnras/stv044}

\bibitem[{{Hirano} {et~al.}(2014){Hirano}, {Hosokawa}, {Yoshida}, {Umeda}, {Omukai}, {Chiaki}, \& {Yorke}}]{Hirano2014}
{Hirano}, S., {Hosokawa}, T., {Yoshida}, N., {et~al.} 2014, \apj, 781, 60, \dodoi{10.1088/0004-637X/781/2/60}

\bibitem[{{Ishigaki} {et~al.}(2018){Ishigaki}, {Tominaga}, {Kobayashi}, \& {Nomoto}}]{Ishigaki2018}
{Ishigaki}, M.~N., {Tominaga}, N., {Kobayashi}, C., \& {Nomoto}, K. 2018, \apj, 857, 46, \dodoi{10.3847/1538-4357/aab3de}

\bibitem[{{Jeena} {et~al.}(2024){Jeena}, {Banerjee}, \& {Heger}}]{Jeena2024}
{Jeena}, S.~K., {Banerjee}, P., \& {Heger}, A. 2024, \mnras, 527, 4790, \dodoi{10.1093/mnras/stad3498}

\bibitem[{{Klessen} \& {Glover}(2023)}]{Klessen2023}
{Klessen}, R.~S., \& {Glover}, S. C.~O. 2023, \araa, 61, 65, \dodoi{10.1146/annurev-astro-071221-053453}

\bibitem[{{Kobayashi}(2012)}]{Kobayashi2012}
{Kobayashi}, C. 2012, in Astronomical Society of the Pacific Conference Series, Vol. 458, Galactic Archaeology: Near-Field Cosmology and the Formation of the Milky Way, ed. W.~{Aoki}, M.~{Ishigaki}, T.~{Suda}, T.~{Tsujimoto}, \& N.~{Arimoto}, 113, \dodoi{10.48550/arXiv.1202.3356}

\bibitem[{{Konstantopoulou} {et~al.}(2024){Konstantopoulou}, {De Cia}, {Ledoux}, {Krogager}, {Mattsson}, {Watson}, {Heintz}, {P{\'e}roux}, {Noterdaeme}, {Andersen}, {Fynbo}, {Jermann}, \& {Ramburuth-Hurt}}]{Konstantopoulou2023}
{Konstantopoulou}, C., {De Cia}, A., {Ledoux}, C., {et~al.} 2024, \aap, 681, A64, \dodoi{10.1051/0004-6361/202347171}

\bibitem[{{Koutsouridou} {et~al.}(2024){Koutsouridou}, {Salvadori}, \& {Sk{\'u}lad{\'o}ttir}}]{Koutsouridou2024}
{Koutsouridou}, I., {Salvadori}, S., \& {Sk{\'u}lad{\'o}ttir}, {\'A}. 2024, \apjl, 962, L26, \dodoi{10.3847/2041-8213/ad2466}

\bibitem[{{Koutsouridou} {et~al.}(2023){Koutsouridou}, {Salvadori}, {Sk{\'u}lad{\'o}ttir}, {Rossi}, {Vanni}, \& {Pagnini}}]{Koutsouridou2023a}
{Koutsouridou}, I., {Salvadori}, S., {Sk{\'u}lad{\'o}ttir}, {\'A}., {et~al.} 2023, \mnras, 525, 190, \dodoi{10.1093/mnras/stad2304}

\bibitem[{{Limongi} \& {Chieffi}(2018)}]{Limongi2018}
{Limongi}, M., \& {Chieffi}, A. 2018, \apjs, 237, 13, \dodoi{10.3847/1538-4365/aacb24}

\bibitem[{{Ma} {et~al.}(2017{\natexlab{a}}){Ma}, {Maio}, {Ciardi}, \& {Salvaterra}}]{Ma2017b}
{Ma}, Q., {Maio}, U., {Ciardi}, B., \& {Salvaterra}, R. 2017{\natexlab{a}}, \mnras, 472, 3532, \dodoi{10.1093/mnras/stx1839}

\bibitem[{{Ma} {et~al.}(2017{\natexlab{b}}){Ma}, {Maio}, {Ciardi}, \& {Salvaterra}}]{Ma2017}
---. 2017{\natexlab{b}}, \mnras, 466, 1140, \dodoi{10.1093/mnras/stw3159}

\bibitem[{{Nomoto} {et~al.}(2013){Nomoto}, {Kobayashi}, \& {Tominaga}}]{Nomoto2013}
{Nomoto}, K., {Kobayashi}, C., \& {Tominaga}, N. 2013, \araa, 51, 457, \dodoi{10.1146/annurev-astro-082812-140956}

\bibitem[{{Pallottini} {et~al.}(2014){Pallottini}, {Ferrara}, {Gallerani}, {Salvadori}, \& {D'Odorico}}]{Pallottini2014}
{Pallottini}, A., {Ferrara}, A., {Gallerani}, S., {Salvadori}, S., \& {D'Odorico}, V. 2014, \mnras, 440, 2498, \dodoi{10.1093/mnras/stu451}

\bibitem[{{Ritter} {et~al.}(2012){Ritter}, {Safranek-Shrader}, {Gnat}, {Milosavljevi{\'c}}, \& {Bromm}}]{Ritter2012}
{Ritter}, J.~S., {Safranek-Shrader}, C., {Gnat}, O., {Milosavljevi{\'c}}, M., \& {Bromm}, V. 2012, \apj, 761, 56, \dodoi{10.1088/0004-637X/761/1/56}

\bibitem[{{Rossi} {et~al.}(2021){Rossi}, {Salvadori}, \& {Sk{\'u}lad{\'o}ttir}}]{Rossi2021}
{Rossi}, M., {Salvadori}, S., \& {Sk{\'u}lad{\'o}ttir}, {\'A}. 2021, \mnras, 503, 6026, \dodoi{10.1093/mnras/stab821}

\bibitem[{{Rossi} {et~al.}(2023){Rossi}, {Salvadori}, {Sk{\'u}lad{\'o}ttir}, \& {Vanni}}]{Rossi2023}
{Rossi}, M., {Salvadori}, S., {Sk{\'u}lad{\'o}ttir}, {\'A}., \& {Vanni}, I. 2023, \mnras, 522, L1, \dodoi{10.1093/mnrasl/slad029}

\bibitem[{{Saccardi} {et~al.}(2023){Saccardi}, {Salvadori}, {D'Odorico}, {Cupani}, {Fumagalli}, {Berg}, {Becker}, {Ellison}, \& {Lopez}}]{Saccardi2023}
{Saccardi}, A., {Salvadori}, S., {D'Odorico}, V., {et~al.} 2023, \apj, 948, 35, \dodoi{10.3847/1538-4357/acc39f}

\bibitem[{{Salvadori} {et~al.}(2019){Salvadori}, {Bonifacio}, {Caffau}, {Korotin}, {Andreevsky}, {Spite}, \& {Sk{\'u}lad{\'o}ttir}}]{Salvadori2019}
{Salvadori}, S., {Bonifacio}, P., {Caffau}, E., {et~al.} 2019, \mnras, 487, 4261, \dodoi{10.1093/mnras/stz1464}

\bibitem[{{Salvadori} {et~al.}(2023){Salvadori}, {D'Odorico}, {Saccardi}, {Sk{\'u}lad{\'o}ttir}, \& {Vanni}}]{Salvadori2023}
{Salvadori}, S., {D'Odorico}, V., {Saccardi}, A., {Sk{\'u}lad{\'o}ttir}, {\'A}., \& {Vanni}, I. 2023, in Memorie della Societa Astronomica Italiana, Vol.~94, 215, \dodoi{10.36116/MEMSAIT_94N2.2023.215}

\bibitem[{{Salvadori} \& {Ferrara}(2012)}]{Salvadori2012}
{Salvadori}, S., \& {Ferrara}, A. 2012, \mnras, 421, L29, \dodoi{10.1111/j.1745-3933.2011.01200.x}

\bibitem[{{Skinner} \& {Wise}(2020)}]{Skinner2020}
{Skinner}, D., \& {Wise}, J.~H. 2020, \mnras, 492, 4386, \dodoi{10.1093/mnras/staa139}

\bibitem[{{Sk{\'u}lad{\'o}ttir} {et~al.}(2024){Sk{\'u}lad{\'o}ttir}, {Vanni}, {Salvadori}, \& {Lucchesi}}]{Skuladottir2023}
{Sk{\'u}lad{\'o}ttir}, {\'A}., {Vanni}, I., {Salvadori}, S., \& {Lucchesi}, R. 2024, \aap, 681, A44, \dodoi{10.1051/0004-6361/202346231}

\bibitem[{{Sk{\'u}lad{\'o}ttir} {et~al.}(2021){Sk{\'u}lad{\'o}ttir}, {Salvadori}, {Amarsi}, {Tolstoy}, {Irwin}, {Hill}, {Jablonka}, {Battaglia}, {Starkenburg}, {Massari}, {Helmi}, \& {Posti}}]{Skuladottir2021}
{Sk{\'u}lad{\'o}ttir}, {\'A}., {Salvadori}, S., {Amarsi}, A.~M., {et~al.} 2021, \apjl, 915, L30, \dodoi{10.3847/2041-8213/ac0dc2}

\bibitem[{{Smith} {et~al.}(2015){Smith}, {Wise}, {O'Shea}, {Norman}, \& {Khochfar}}]{Smith2015}
{Smith}, B.~D., {Wise}, J.~H., {O'Shea}, B.~W., {Norman}, M.~L., \& {Khochfar}, S. 2015, \mnras, 452, 2822, \dodoi{10.1093/mnras/stv1509}

\bibitem[{{Susa} {et~al.}(2014){Susa}, {Hasegawa}, \& {Tominaga}}]{Susa2014}
{Susa}, H., {Hasegawa}, K., \& {Tominaga}, N. 2014, \apj, 792, 32, \dodoi{10.1088/0004-637X/792/1/32}

\bibitem[{{Takahashi}(2018)}]{Takahashi2018b}
{Takahashi}, K. 2018, \apj, 863, 153, \dodoi{10.3847/1538-4357/aad2d2}

\bibitem[{{Takahashi} {et~al.}(2018){Takahashi}, {Yoshida}, \& {Umeda}}]{Takahashi2018}
{Takahashi}, K., {Yoshida}, T., \& {Umeda}, H. 2018, \apj, 857, 111, \dodoi{10.3847/1538-4357/aab95f}

\bibitem[{{Vanni} {et~al.}(2023{\natexlab{a}}){Vanni}, {Salvadori}, \& {Sk{\'u}lad{\'o}ttir}}]{vanni2023}
{Vanni}, I., {Salvadori}, S., \& {Sk{\'u}lad{\'o}ttir}, {\'A}. 2023{\natexlab{a}}, in Memorie della Societa Astronomica Italiana, Vol.~94, 84, \dodoi{10.36116/MEMSAIT_94N2.2023.84}

\bibitem[{{Vanni} {et~al.}(2023{\natexlab{b}}){Vanni}, {Salvadori}, {Sk{\'u}lad{\'o}ttir}, {Rossi}, \& {Koutsouridou}}]{Vanni2023b}
{Vanni}, I., {Salvadori}, S., {Sk{\'u}lad{\'o}ttir}, {\'A}., {Rossi}, M., \& {Koutsouridou}, I. 2023{\natexlab{b}}, \mnras, 526, 2620, \dodoi{10.1093/mnras/stad2910}

\bibitem[{{Vladilo} {et~al.}(2018){Vladilo}, {Gioannini}, {Matteucci}, \& {Palla}}]{Vladilo2018}
{Vladilo}, G., {Gioannini}, L., {Matteucci}, F., \& {Palla}, M. 2018, \apj, 868, 127, \dodoi{10.3847/1538-4357/aae8dc}

\bibitem[{{Welsh} {et~al.}(2019){Welsh}, {Cooke}, \& {Fumagalli}}]{Welsh2019}
{Welsh}, L., {Cooke}, R., \& {Fumagalli}, M. 2019, \mnras, 487, 3363, \dodoi{10.1093/mnras/stz1526}

\bibitem[{{Welsh} {et~al.}(2022){Welsh}, {Cooke}, {Fumagalli}, \& {Pettini}}]{Welsh2022}
{Welsh}, L., {Cooke}, R., {Fumagalli}, M., \& {Pettini}, M. 2022, \apj, 929, 158, \dodoi{10.3847/1538-4357/ac4503}

\bibitem[{Xing {et~al.}(2023)Xing, Zhao, Liu, Heger, Han, Aoki, Chen, Ishigaki, Li, \& Zhao}]{Xing2023}
Xing, Q.-F., Zhao, G., Liu, Z.-W., {et~al.} 2023, Nature, \dodoi{10.1038/s41586-023-06028-1}

\end{thebibliography}
\bibliographystyle{aasjournal}

 \appendix 

 \section{Analysis of X-Shooter quasar spectra}
 \label{app:abs_sys}
In this section, we report the results of the fit of three absorption systems discussed in \citet{Christensen2023} for which we analysed the corresponding X-Shooter spectra. The spectrum of UHS J0439+1634 is the one released by the XQR-30 collaboration \citep{Dodorico2023}, while the 
spectrum of ULAS J1342+0928 was reduced and analysed in this work. Briefly, ULAS J1342+0928 was observed with VLT/X-Shooter in 2017 and 2018 (PID 098.B-0537, 0100.A-0898) for $\sim 23$h adopting a slit of 0.6" in the NIR arm. We retrieved the raw frames from the ESO archive and reduced them with the custom pipeline described in \citet[][]{Dodorico2023}. The combined spectrum has a nominal resolving power $R\simeq 8100$ and a signal-to-noise ratio S/N~$\simeq 19$ per 10 km s$^{-1}$ bin at $\lambda=1285$ $\mathring{\rm A}$ rest frame. 
This value, converted to the JWST/NIRSpec wavelength bin at $\lambda \sim1.1$ $\mu$m ($\Delta v \simeq 64$ km s$^{-1}$) increases to S/N~$\simeq 50$, comparable to the one  of the spectra analysed by \citet[][S/N $\approx 60$, see their Fig.~5]{Christensen2023}.

The absorption lines have been fitted with Voigt profiles using the Python package Astrocook \citep{Cupani2022}. 
For the three analysed systems, we assumed that the Doppler parameters were the same for all observed transitions. In the case of the two systems along the spectrum of ULAS J1342+0928, we fixed the Doppler parameter to the minimum value allowed by the resolving power of the observations 
(see Sodini et al. subm.). However, we checked that decreasing the Doppler parameter to 6 km s$^{-1}$ the results do not change significantly. Due to the limited NIR coverage of the X-Shooter spectrum with respect to the NIRSpec one, we do not cover the \MGII\ doublet transitions for either system. 

\begin{figure}
    \centering
    \includegraphics[width=8cm]{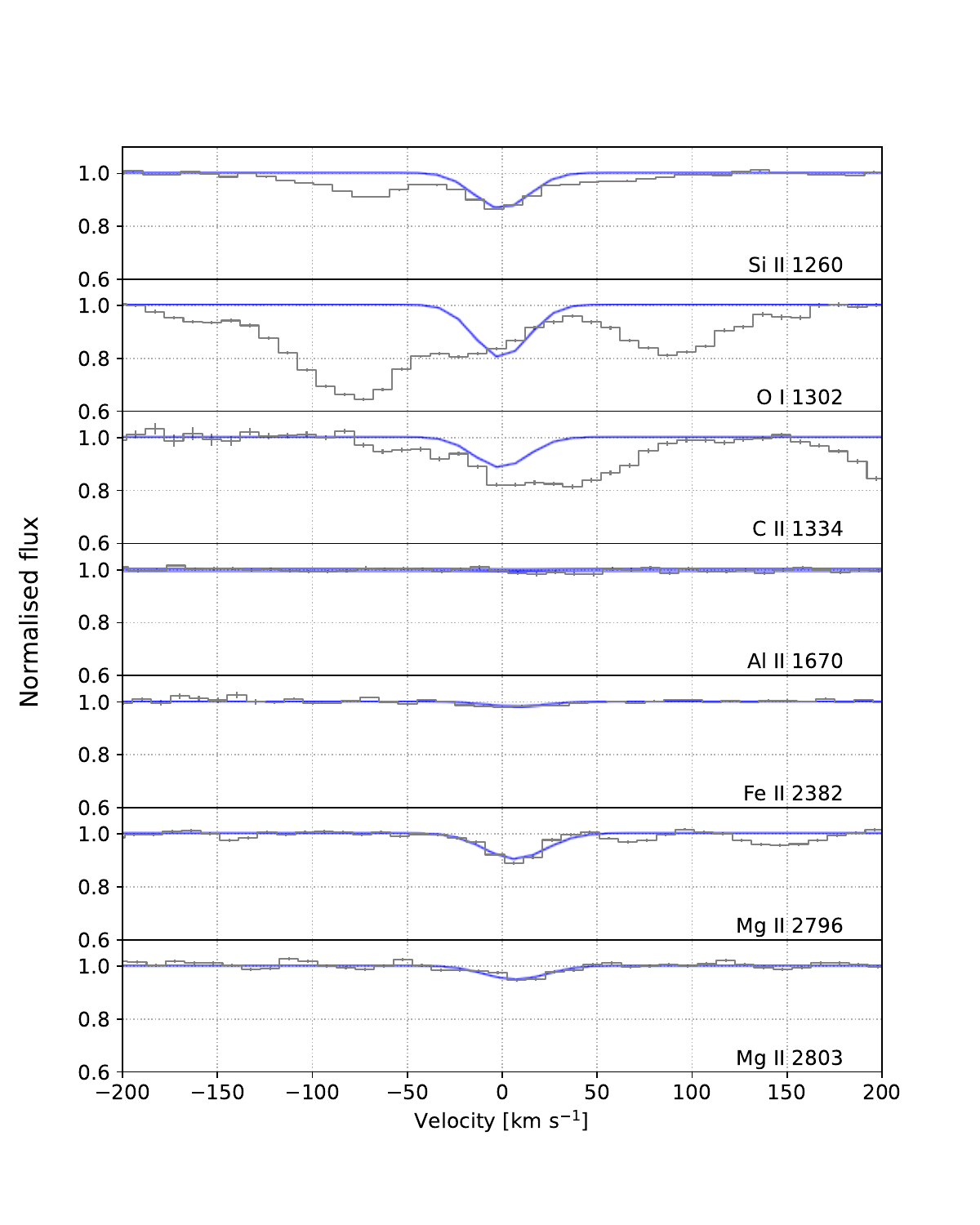}
    \caption{Transitions detected at $z=6.2897$ in the X-Shooter spectrum of UHS J0439+1634. The observed flux is plotted as a grey histogram with 1$\sigma$ errorbars plotted for each pixel, the Voigt profile model is plotted as a blue solid line. This figure can be compared with Figure B.45 in \citet{Christensen2023}, noting the different velocity scale. }
    \label{fig:sys6289}
\end{figure}

\begin{figure}
    \centering
    \includegraphics[width=8cm]{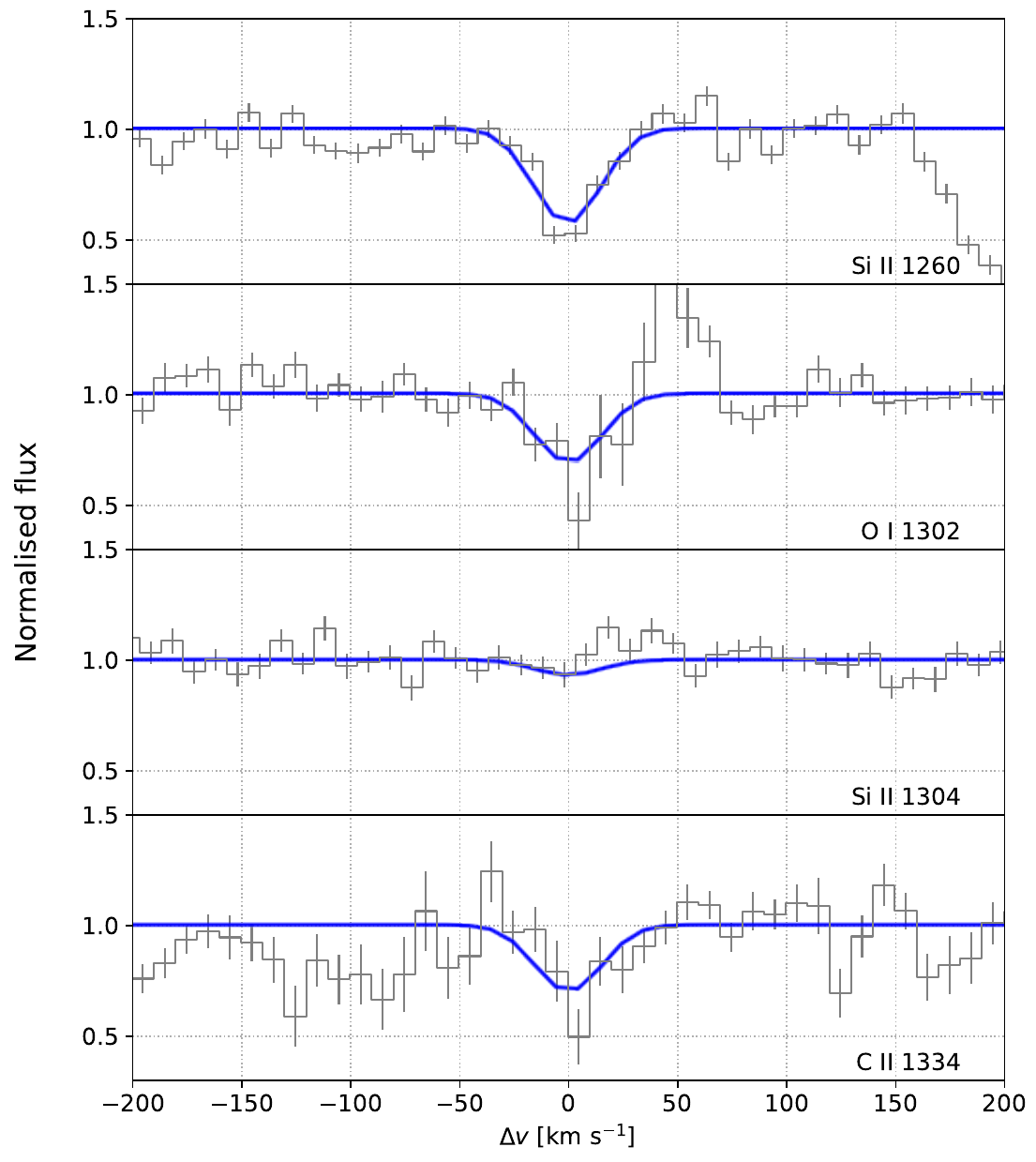}
    \caption{Transitions detected at $z=7.368$ in the X-Shooter spectrum of ULAS J1342+0928. The lines are the same as in Fig.~\ref{fig:sys6289}. This figure can be compared with Figure B.59 in \citet{Christensen2023}, noting the different velocity scale. }
    \label{fig:sysz7368}
\end{figure}

\begin{figure}
    \centering
    \includegraphics[width=8cm]{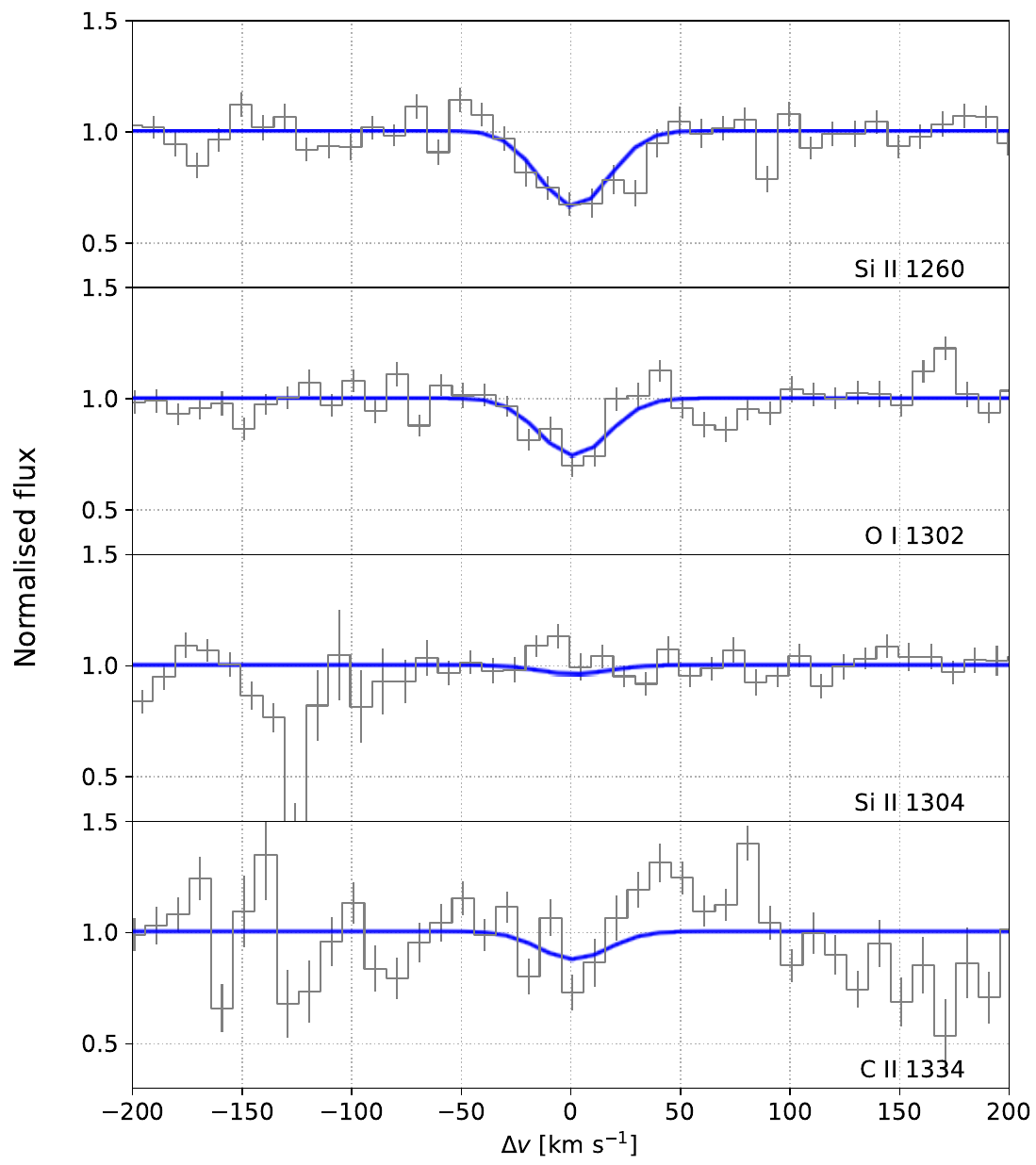}
    \caption{Transitions detected at $z=7.4434$ in the X-Shooter spectrum of ULAS J1342+0928. The lines are the same as in Fig.~\ref{fig:sys6289}. This figure can be compared with Figure B.60 in \citet{Christensen2023}, noting the different velocity scale.}
    \label{fig:sysz7443}
\end{figure}

In the case of the system at $z=6.2897$ in the spectrum of UHS J0439+1634, \citet{Christensen2023} identify as \OIa\ 1302 $\mathring{\rm A}$ 
at $z=6.288$ an absorption feature which  \citet{RDavies2023a} fit in the X-Shooter spectrum as a \CIV\ doublet at $z=5.1299$. 
For the purpose of this paper and in order to compare with the results by \citet{Christensen2023}, we derived a conservative upper limit on the 
\OIa\ column density fitting the observed absorption assuming the redshift and the Doppler parameter of the \SII\ 1260 $\mathring{\rm A}$ absorption 
line of this system, i.e. assuming $z\approx 6.2897$. 
The very good quality of the X-Shooter spectrum allowed us also to measure the column density of a very weak \FeII\ 2382 $\mathring{\rm A}$ absorption line and to determine an upper limit to the transition of \AlII\ 1670 $\mathring{\rm A}$.
The absorption lines and the Voigt profile fit for the three studied systems are shown in Fig.~\ref{fig:sys6289}, \ref{fig:sysz7368} and \ref{fig:sysz7443}; while the measured column densities are reported in Table~\ref{tab:fit}.

\begin{table*}[t]
    \centering
    \begin{tabular}{c c c c }
    \hline
    QSO & UHS J0439+1634 & ULAS J1342+0928 & ULAS J1342+0928 \\         
    $z_{\rm abs}$ & $6.2897$ & $7.36899$ & $7.44345$ \\
    $b$ [km s$^{-1}$] & $16\pm2$ & 8 & 8 \\ 
    \hline
    C II & $13.03\pm0.08$ & $13.57\pm0.23$ & $13.07\pm0.33$ \\
    O I & $<13.86$  & $14.03\pm0.19$ & $13.91\pm0.1$ \\
    Si II & $12.22\pm0.03$ & $12.95\pm0.10$ & $12.70\pm0.10$ \\
    Mg II  & $11.97\pm0.05$ & -- & -- \\
    Al II & $<10.33$ & -- & -- \\
    Fe II  & $11.55\pm0.04$ & $<12.70$ &  $<13.10$ \\
    \hline
\end{tabular}
    \caption{Column densities for the systems reported by  \citet[][]{Christensen2023} along the sightlines to UHS J0439+1634 and ULAS J1342+0928 determined in the X-Shooter spectra.}
    \label{tab:fit}
\end{table*}

\section{Environments enriched by a PISN and another Pop~III SN}
\label{app:twopopIII}

Cosmological simulations predict that the first stars formed preferentially isolated, but they could also appear in small groups \citep[see e.g.][]{Skinner2020}. Furthermore, 
in this work we aim at providing chemical diagnostics to study the fingerprints of the first stars in distant absorbers, including Lyman Limit systems and sub-DLAs. The nature of these systems is still unclear and they might be associated to gaseous filaments enriched by different low-mass galaxies hosting Pop~III stars. For these reasons, in this Section we investigate how the chemical diagnostics for PISN-enriched environments are affected when considering the contribution of a second Pop~III SN, i.e. further exploring the case $f_{\rm PopIII}=1$.

\begin{figure*}
    \includegraphics[width=1\linewidth]{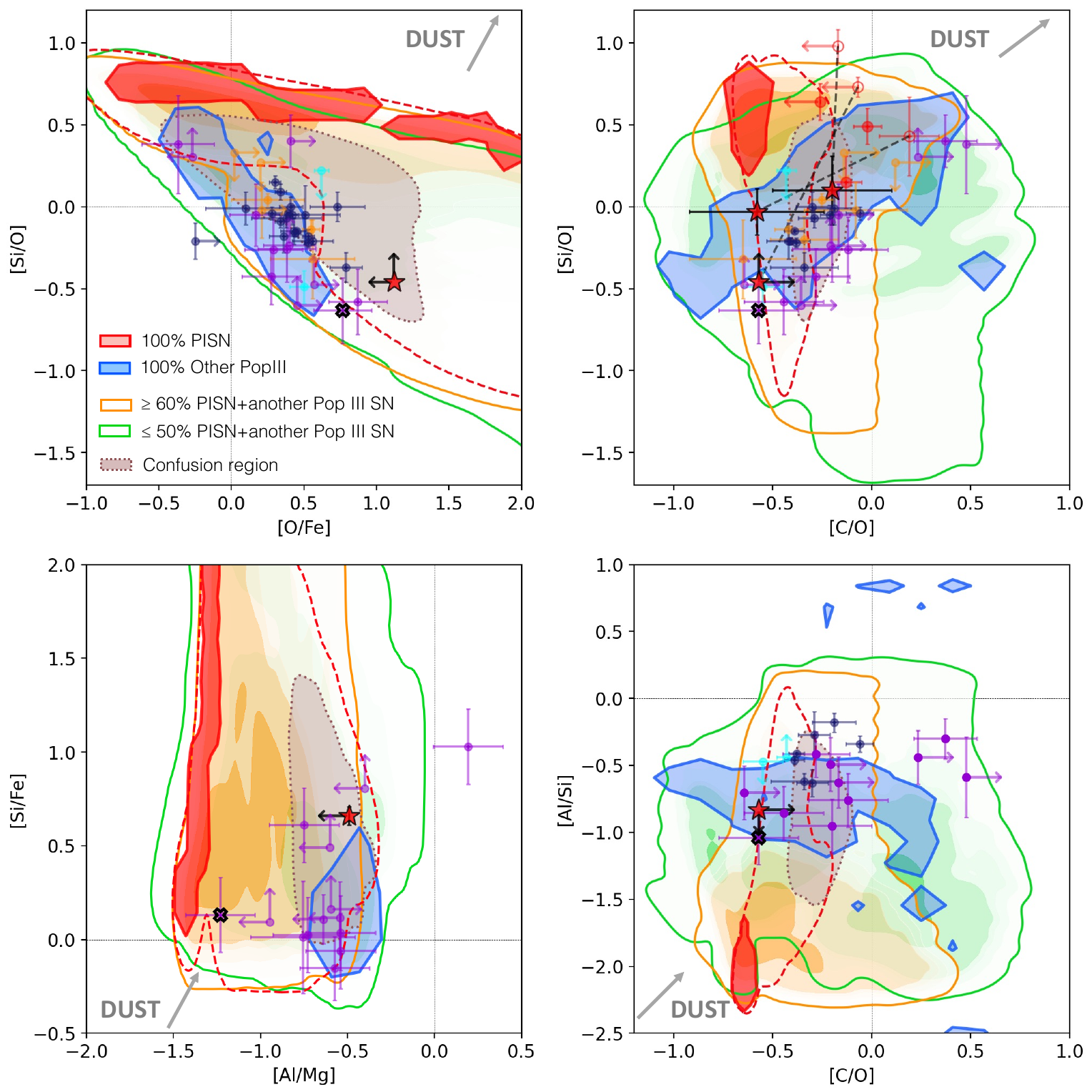}
    \caption{Same as Fig.~\ref{fig:Maps_DLAs}, but the chemical abundances are inherited by: only PISNe (solid red), only other Pop~III SNe (solid blue), a single PISN but also another Pop~III SN in different proportions ($\geq 60\%$ PISN: solid orange, $\leq 50\%$ PISN: solid green). We show also the chemical abundances predicted for environments enriched mainly by PISNe but also by Pop~II SNe ($f_{\rm PopIII} \geq 0.6$, dashed red) and mainly by Pop~II SNe but also by PISNe ($f_{\rm PopIII} \leq 0.5$, dotted grey, see Figs.~\ref{fig:Maps_DLAs}-\ref{fig:Maps_PISN}).}
    \label{fig:two_PopIII_1}
\end{figure*}

\begin{figure*}
    \includegraphics[width=1\linewidth]{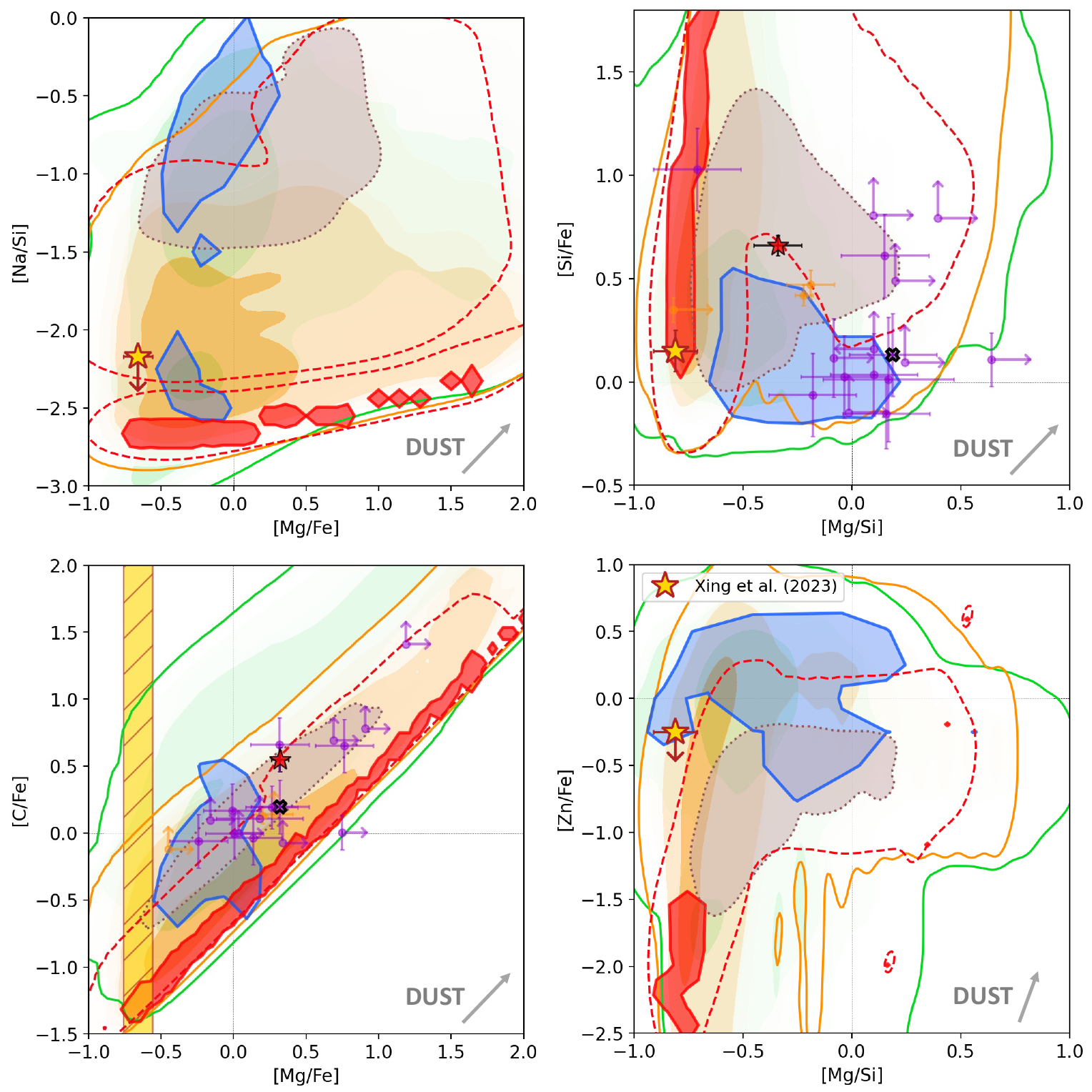}
    \caption{The same as Fig.~\ref{fig:Maps_PISN}, but with the additional models described in the caption of Fig.~\ref{fig:two_PopIII_1}.}
    \label{fig:two_PopIII_2}
\end{figure*}

The results of our investigation are shown in Figs.~\ref{fig:two_PopIII_1}-\ref{fig:two_PopIII_2}, where we compare the chemical abundance ratios predicted for environments enriched by one PISN and another Pop~III SN (i.e. two Pop~III SNe and $f_{\rm PopIII} = 1$) with those previously shown in Figs.~\ref{fig:Maps_DLAs}-\ref{fig:Maps_PISN}.
Clearly, these abundance ratios are widespread in our chemical diagnostics, covering the areas predicted for environments enriched by single PISNe, other Pop~III SNe, and a mixture of PISNe and Pop~II SN, i.e. the so-called confusion regions.
Thus, absorbers in the confusions regions might be enriched by Pop~III only ($f_{\rm PopIII}=1$) through a combination of a PISN and another Pop~III SN, or predominantly enriched by a populations of normal Pop~II SNe ($f_{\rm PopIII}\leq 0.5$). To uncover the nature of these absorbers we should use all the proposed diagnostics and, eventually, use a $\chi^2$ fitting procedure (see Appendix~\ref{app:chisq}) exploiting their overall abundance pattern.

It is important to note that {\it in all our diagnostics} the absorber J1018+0548 (black cross) is consistent for being enriched by a PISN and another Pop~III SN, i.e., uniquely imprinted by Pop~III stars. To unveil the properties of their progenitors, we need to compare its observed chemical abundance pattern with our theoretical predictions as we will explain in the next Section.

\section{Fitting procedure}
\label{app:chisq}

The $\chi^2$ is computed following the equation proposed by \citet{Heger2010}, which is general and allows us to include the contribution of upper/lower limit measurements in the calculation:
\begin{equation}
    \chi^2=\sum_{i=1}^N \dfrac{(O_i-T_i)^2}{\sigma^2_{O_i}+\sigma^2_{T_i}}+\sum_{i=N+1}^{N+U} \dfrac{(O_i-T_i)^2}{\sigma^2_{T_i}}\times\Theta(O_i-T_i)+\sum_{i=N+U+1}^{N+U+L}\dfrac{(O_i-T_i)^2}{\sigma^2_{T_i}}\times\Theta(T_i-O_i),
\end{equation}
where $O_i$ and $T_i$ are the observed and theoretically predicted abundance ratios [X$_i$/Fe]
and $\sigma_{O_i, T_i}$ are their respective errors. In order to account for the uncertainties on the stellar yields, we assume $\sigma_{T_i}=0.50$, as suggested by \citet{Nomoto2013} and \citet{Hartwig2018}. 
$\Theta(x)$ is the Heaviside function.
The reduced chi-square, $\chi^2_{\nu}$, is computed as $\chi^2/(N+L+U-M)$, where N is the number of data points with finite measurements, L (U) is the number of data points with lower (upper) limits and M is the number of free parameters of the model (M$=4$ with fixed mixing parameter of Pop III SNe). Finally, we define the relative chi square as $\chi^2_{rel} \coloneqq \chi^2_{\nu}/min(\chi^2_{\nu})$.

To find the best model for the enrichment of the absorption system at $z = 3.39$ in the sightline of the QSO J1018+0548 \citep{Saccardi2023}, we look for the minimum $\chi^2_{\nu}$ 
by considering all the models presented in Sec.~\ref{sec:model} 
(see Fig.~\ref{fig:PopIII_only}, \ref{fig:Maps_DLAs}, \ref{fig:Maps_PISN}), i.e. by accounting for an enrichment: (i) uniquely driven by a single PISN or a Pop~III SN with different mass/energy; (ii) mainly ($\geq 60\%$ metals) driven by a single PISN/other Pop~III SN with the contribution of Pop~II SNe; (iii) only partially ($\leq 50\%$ metals) driven by a single PISN/other Pop~III SN with the contribution of Pop~II SNe. Furthermore, we also consider an imprint from two Pop~III SNe with various masses and energies, including PISNe, with $f_{\rm PopIII}=1$ (i.e. excluding the contribution of Pop~II SNe). As shown in Fig.~\ref{fig:chi_square} these models provide the best fits for the abundance pattern of J1018+0548.

\begin{table*}
\centering
\begin{tabular}{c|c|c|c|c|c|c}
    \hline
     $\chi^2_{\nu}$ & $f_{\rm PopIII}$ & $M_1 \, \rm [M_{\odot}]$ & $E_1 \, \rm [10^{51} \, erg]$ & $ M_2 \, \rm [M_{\odot}]$ & $ E_2 \, \rm [10^{51} \, erg]$ & $f\rm _1$ \\
     \hline 
     0.19 & 1.0 & 100.0 & 1.8 & 11.8 & 1.8 & 0.8 \\
     0.37 & 1.0 & 253.0 & 97.3 & 27.5 & 1.2 & 0.1 \\
     \hline

\end{tabular}
    \caption{Parameters of the best-fitting models of J1018+0548 shown in Fig.~\ref{fig:chi_square}. $M_{1,2}$ and $E_{1,2}$ are the initial masses and SN explosion energies of the two Pop~III stars polluting the environment. We indicate with $f_1$ the fraction of metals provided by the Pop~III SN that explodes first. Thus, the fraction of metals contributed by the second SN can be computed as $f_2=1-f_1$.}
    \label{tab:bestmod}
\end{table*}

\begin{figure}
    \centering
    \includegraphics[width=0.8\linewidth]{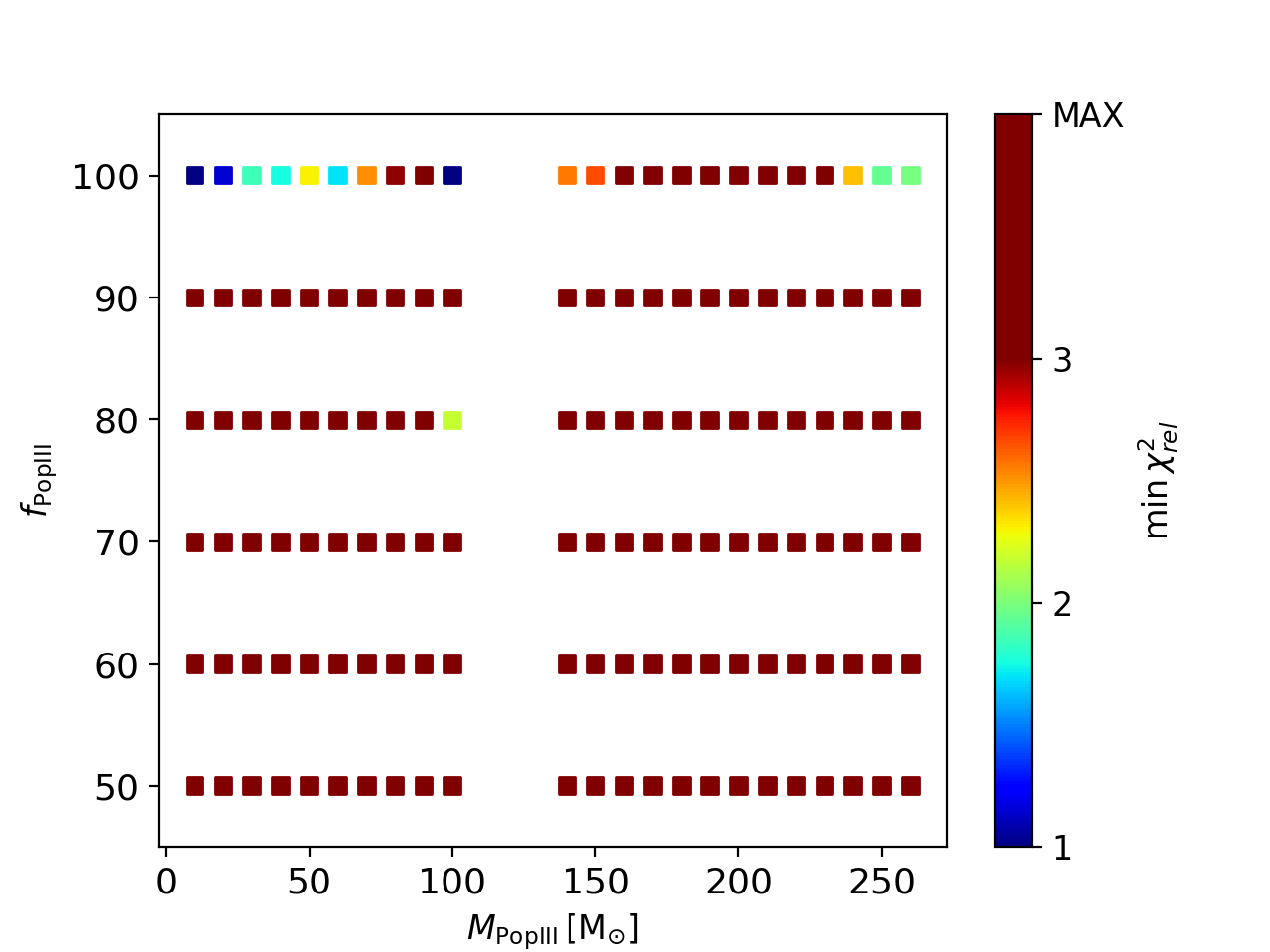}
    \caption{Minimum relative reduced chi square computed fitting the abundances of J1018+0548 for models with different $f_{\rm PopIII}$ and different initial masses of the SN that explodes first, $M_{\rm PopIII}$ or $M_1$.}
    \label{fig:map_chisquare}
\end{figure}

In Table~\ref{tab:bestmod}, we present the parameters of the models displayed in Fig.~\ref{fig:chi_square}, which correspond to an enrichment driven by: 
(i) two Pop~III SNe with same explosion energy, $E_{\rm 1,2}=1.8 \times 10^{51}$~erg, but very different masses, $M_1=100 \rm M_{\odot}$ and $M_2=11.8 \rm M_{\odot}$ ($\chi^2_{\nu}=0.19$);
(ii) a very massive PISN of $M_1=253 \rm M_{\odot}$ and a second Pop~III SN with lower mass, $M_2=27.5 \rm M_{\odot}$, and lower explosion energy, $E_{\rm 2}=1.2 \times 10^{51}$~erg ($\chi^2_{\nu}=0.37$). 
The parameter $f_1$ in the Table represents the fraction of metals provided by the Pop~III SN/PISN that explodes first, while the fraction contributed by the second one will be $f_2=1-f_1$. 

In Fig.~\ref{fig:map_chisquare} we show the minimum $\chi^2_{rel}$ obtained by studying all models with different $f_{\rm PopIII}$ and mass of the SN that explodes first, $M_{\rm PopIII}$, regardless of all the other parameters, i.e. $E_1, M_2, E_2, f_1, M_{\rm PopII}$. The chi square shown for models with $f_{\rm PopIII}=1$ can be either for environments enriched by single Pop~III SNe with $M=M_{\rm PopIII}$ or multiple Pop~III SNe, where the first SN that explodes has $M_1=M_{\rm PopIII}$.

We can clearly see in the Figure that all models with $f_{\rm PopIII} < 1$ have $\chi^2_{rel} > 3$, apart from a specific model with $f_{\rm PopIII}=0.8$ (providing $\chi^2_{rel}=2.21$, $\chi^2=0.45$) that describes the combined pollution by a single Pop~III SN with $M = \rm 100 M_{\odot}$ and a normal core-collapse Pop~II SN with $M = \rm 13 M_{\odot}$. Thus, we can safely exclude that the absorber J1018+0548 has been predominantly imprinted by Pop~II SNe. As shown in the Figure, the models with the lowest $\chi^2_{rel}$ are the ones with $f_{\rm PopIII}=1$, i.e., those uniquely polluted by Pop~III SNe: both PISNe+Pop III SNe and two Pop~III SNe. We chose to show in Fig.~\ref{fig:chi_square} the abundance patterns of the best models of these two categories, i.e. models that have $\chi^2_{rel} \leq 2$ and that account for an enrichment by two Pop~III SNe or by one PISN and a Pop~III SN.

\end{document}